\documentclass[preprint]{aastex61}
\usepackage{xspace}
\usepackage{soul}
\usepackage{color}

\newcommand{\Fermi}{\emph{Fermi}\xspace}

\newcommand{\tgw}{t$_{\rm GW}$\xspace}
\newcommand{\tem}{t$_{\rm EM}$\xspace}

\shorttitle{\Fermi-LAT observations of the LIGO/Virgo event GW170817}
\shortauthors{Fermi-LAT team}

\begin{document}

\title{\Fermi-LAT observations of the LIGO/Virgo event GW170817}

\author{the \Fermi-LAT Collaboration}
\correspondingauthor{Daniel Kocevski (daniel.kocevski@nasa.gov), Nicola Omodei (nicola.omodei@stanford.edu), Giacomo Vianello (giacomov@stanford.edu)}

\begin{abstract}
We present the \Fermi Large Area Telescope (LAT) observations of the binary neutron star merger event GW170817 and the associated short gamma-ray burst (SGRB) GRB\,170817A detected by the \Fermi Gamma-ray Burst Monitor. The LAT was entering the South Atlantic Anomaly at the time of the LIGO/Virgo trigger ($t_{\rm GW}$) and therefore cannot place constraints on the existence of high-energy (E $>$ 100 MeV) emission associated with the moment of binary coalescence. We focus instead on constraining high-energy emission on longer timescales. No candidate electromagnetic counterpart was detected by the LAT on timescales of minutes, hours, or days after the LIGO/Virgo detection. The resulting flux upper bound (at 95\% C.L.\/) from the LAT is $4.5\times$10$^{-10}$ erg cm$^{-2}$ s$^{-1}$ in the 0.1--1 GeV range covering a period from \tgw + 1153 s to \tgw + 2027 s. At the distance of GRB\,170817A, this flux upper bound corresponds to a luminosity upper bound of 9.7$\times10^{43}$ erg s$^{-1}$, which is 5 orders of magnitude less luminous than the only other LAT SGRB with known redshift, GRB\,090510. We also discuss the prospects for LAT detection of electromagnetic counterparts to future gravitational wave events from Advanced LIGO/Virgo in the context of GW170817/GRB\,170817A.
\end{abstract}
\keywords{gravitational waves, gamma rays:  general, methods: observation}

\section{Introduction} \label{sec:intro}

Short gamma-ray bursts (SGRBs) have long been thought to be associated with the coalescence of binary compact objects such as members of neutron star-black hole (NS-BH) and neutron star-neutron star (NS-NS) systems~\citep{Paczynski1986, Eichler1989, Paczynski1991, Narayan1992}. This connection arises from considerations of their duration~\citep{Norris1984, Dezalay1991, Kouveliotou1993}, redshift~\citep{Berger2014}, and host galaxy distributions~\citep{2008MNRAS.385L..10T,2009A&A...498..711D,Fong2013}.  The observed absence of associated conventional supernovae~\citep{Fox2005, Hjorth2005a, Hjorth2005b, Soderberg2006, kocevski2010, Berger2013,2016ApJ...827..102T} has further supported this paradigm.  Deep HST observations of nearby SGRBs has also provided tantalizing evidence for their association to kilonovae \citep{Metzger2010, 2012grb..book.....K,tlf+2013,Yang2015NatCo...6E7323Y,Jin2016NatCo...712898J},
short-lived infrared transients powered by the radioactive energy of the NS merger ejecta.

The merging of neutron-star binaries is predicted to also result in emission of gravitational waves \citep[e.g.,][and references therein]{Kobayashi2003}, making SGRBs promising candidates for joint gravitational wave (GW) and electromagnetic (EM) detections. For example, the observed orbital decay rate of the binary pulsar PSR~B1913+16 is consistent with the predictions of General Relativity to better than  10$^{-3}$ precision, setting the date of their merging as $\sim$300 million years from now \citep{1979Natur.277..437T}. 

Confirmation of the long suspected connection between SGRBs and compact binary coalescence was provided on 2017 August 17 when the Advanced Laser Interferometer Gravitational-Wave Observatory (LIGO)\citep{AdvLIGO} and Advanced Virgo experiments \citep{AdvVirgo} triggered on a compact binary merger candidate~\citep{GW170817} coincident in time with a \Fermi Gamma-ray Burst Monitor (GBM) detected SGRB, GRB\,170817A~\citep{GRB170817A_Discovery, Goldstein2017}, which was also detected by INTEGRAL SPI-ACS \citep{GCN21507,Savcheno2017}. The joint GW-EM detection has provided the first compelling observational evidence of the relation of SGRBs to neutron star binary coalescence events and has ushered in an exciting era of multi-messenger astronomy. Since gravitational waves probe the system binding energy and gamma-ray bursts are driven by the configuration of matter outside the ultimate event horizon, the combination of such information opens up discovery space concerning tidal disruption of infalling neutron stars, and associated accretion disk and jet formation. It perhaps also offers the prospect for measuring the mass to radius ratio and equation of state of a participating neutron star prior to destruction. Shortly after the trigger the detection of an optical counterpart in the elliptical galaxy NGC 4993 was announced ~\citep{Coulter2017,GCN21529,GW170817-MMAD}, at the position (R.A., Dec = 197$\fdg$450354, $-$23$\fdg$381484, J2000). We will assume these coordinates as the location of GW170817 for the remainder of the paper.

The \Fermi Large Area Telescope (LAT)  unfortunately was not collecting data at the time of the LIGO/Virgo and GBM triggers due to a passage through the South Atlantic Anomaly (SAA), and was thus unable to observe the prompt emission phase of the GRB. The LAT resumed collecting science data $\sim 10^3$ seconds later, so we focus instead on constraining high-energy emission on longer time scales. We use these limits and the estimated distance of 42.5 Mpc to NGC 4993 
to set upper limits on the luminosity of the late-time emission of GRB\,170817A above 100 MeV. 

We describe the details of the data analysis in $\S$2, compare GRB\,170817A to other LAT-detected SGRBs in $\S$3, discuss the prospects of detecting SGRBs associated with future GW triggers in $\S$4, and conclude in $\S$5.

\section{Analysis}
\subsection{LAT observations of GW170817}

The \Fermi Gamma-ray Space Telescope consists of two primary science instruments, the GBM~\citep{Meegan2009} and the LAT~\citep{Atwood09}. The GBM comprises 14 scintillation detectors designed to study the gamma-ray sky in the energy band of $\sim$8 keV--40 MeV. The LAT is a pair conversion telescope consisting of a 4$\times$4 array of silicon strip trackers and tungsten converters together with Cesium Iodide (CsI) calorimeters covered by a segmented anticoincidence detector to reject charged-particle background events. The LAT is sensitive in an energy range covering 20~MeV to more than 300~GeV with a field of view (FOV) of 2.4~sr, observing the entire sky every two orbits ($\sim$3~hours) by rocking north and south about the orbital plane on alternate orbits~\citep{Atwood09}. The LAT detects roughly 15 GRBs per year above 100 MeV, of which $\sim1-2$ are SGRBs, with localization precisions of $\sim$10 arcmin~\citep{Vianello15}. The high-energy emission associated with SGRBs is substantially longer in duration with respect to their keV--MeV emission as observed by the GBM, having been detected on timescales of $>100$~s in two burst sources, and has been proposed to be related to the afterglow phase of the burst~\citep{Kumar2009, Ackermann2013, Ackermann2014, Kouveliotou2013}. The LAT is currently the only instrument that has detected and localized long-lived high-energy emission from SGRBs, and can substantially reduce the localization uncertainties with respect to GBM, aiding follow-up at other wavelengths.

\begin{figure}
	\begin{center}
     \includegraphics[width=1.0\textwidth]{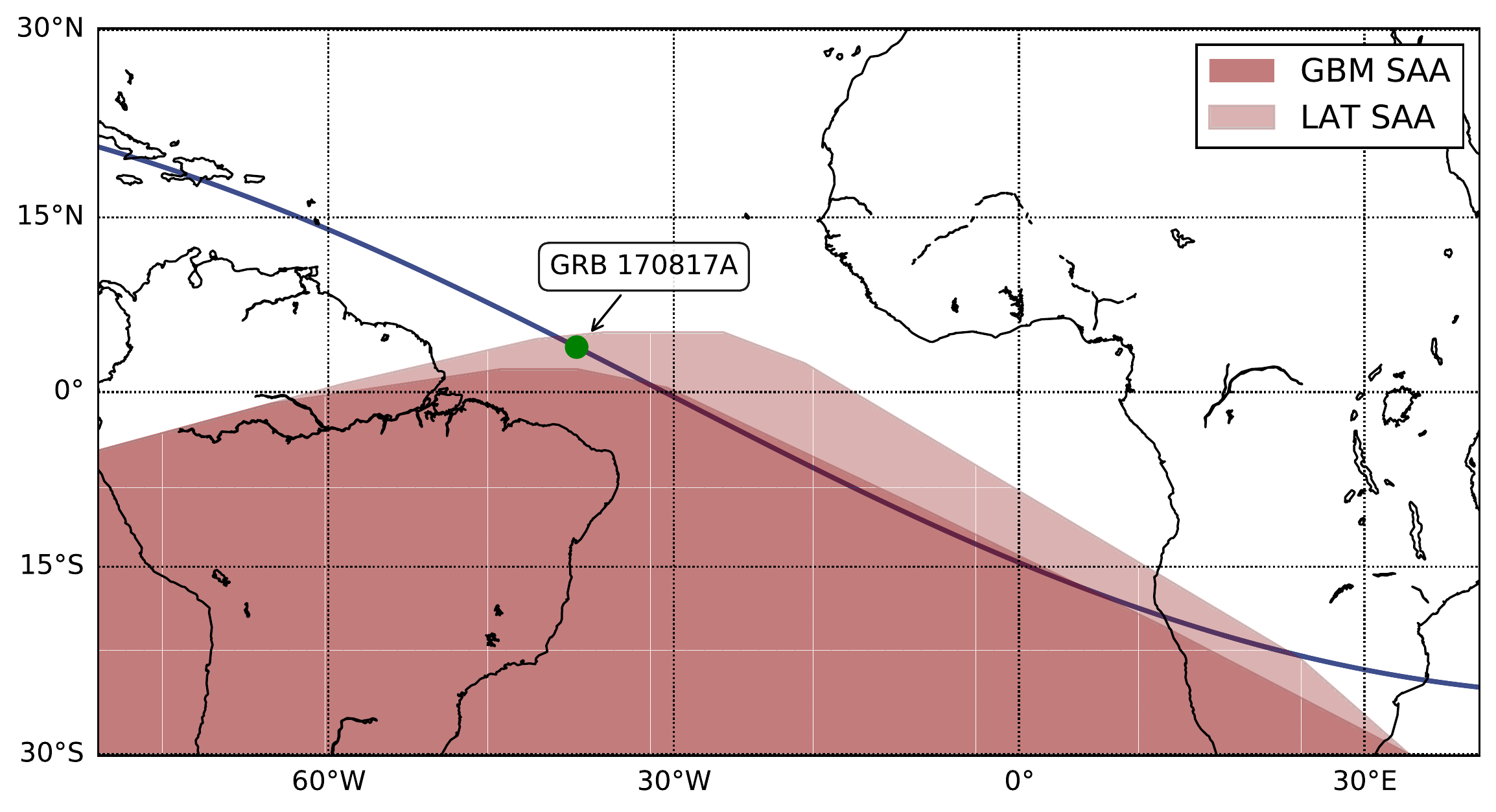}	
	\end{center}
\caption{The position of \Fermi at the trigger time of GRB\,170817A (green dot) and its orbital path from West to East.  The dark and light red regions define the boundaries of the SAA for the GBM and LAT instruments respectively. Both instruments do not collect data inside their respective SAA boundaries due to an elevated charged particle background.}
\label{fig:SAA}
\end{figure}

\begin{figure}
	\begin{center}
     \includegraphics[width=1.0\textwidth]{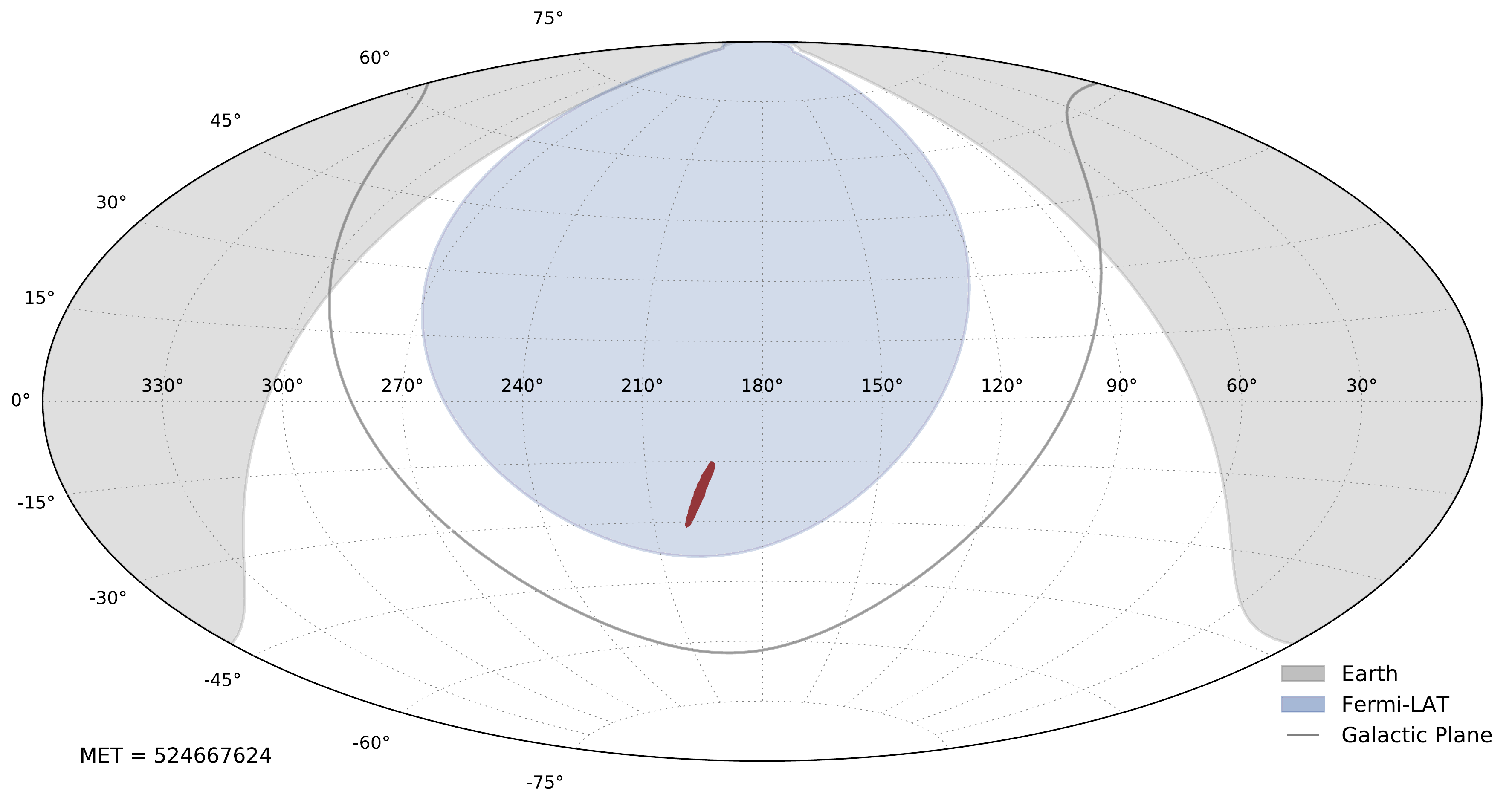}	
	\end{center}
\caption{Location of the LAT field of view upon exiting the SAA at \tgw + 1153 s. The blue and gray regions represent the LAT field of view and the Earth-occulted sky, respectively. The gray line is the Galactic plane. The red region represents the final LIGO/Virgo \texttt{LALInference} map (90\% credible region). The projection is in celestial coordinates.}
\label{fig:Pointing}
\end{figure}

\Fermi-LAT was entering the SAA at the time of the LIGO/Virgo trigger (\tgw = 2017-08-17 12:41:04.444 UTC). During SAA passages the LAT and the GBM do not collect data due to the high charged particle background in this region. Because of the higher susceptibility of the LAT to the charged particles in this region, the SAA boundary employed by the LAT encompasses a $\sim$14\% larger area than the boundary used by the GBM, resulting in slightly different times at which the two instruments do not collect data. The GBM and LAT SAA boundaries are illustrated in Figure~\ref{fig:SAA}. 
At the time of the GBM trigger (t$_{\rm EM}$ = 2017-08-17 12:41:06.474598 UTC), the centroid of the final LIGO/Virgo \texttt{LALInference} map~\citep{veitch2015parameter, gcnLVCGW170817_LAL} using data from all three gravitational-wave observatories (H1, L1, and V1) was located at R.A.=197$\fdg$25, Dec.=$-$25$\fdg$62 (J2000), or Galactic $l$=307$\fdg$9, $b$=37$\fdg$1.  This position was $\theta \sim 90^{\circ}$ from the LAT boresight and outside the nominal $\theta < 65^{\circ}$ LAT FOV. The LAT resumed data taking upon exiting the SAA at t$_{\rm GW}$ + 1153 s. At that time, the entire 90\% credible region of the \texttt{LALInference} map was within the LAT FOV and the region subsequently exited at \tgw + 2027 s. Fig.~\ref{fig:Pointing} shows the sky coverage of the LAT at t$_{\rm GW}$ + 1153 s, when the entire localization region was observed. 

\subsection{Constraints on the high-energy flux of GRB\,170817A}

We searched the LAT data for a gamma-ray counterpart on different time scales before and after the trigger time, and we computed upper bounds on its flux using an unbinned likelihood analysis described in further detail in \citet{Ackermann2016}, \citet{Racusin2017}, and \citet{Vianello2017}. For all the analyses presented here we used the \texttt{P8\_TRANSIENT010E\_V6} events class and the corresponding instrument response functions, and the \Fermi Science Tools version v10r0p5\footnote{http://fermi.gsfc.nasa.gov/ssc/}. We furthermore assume a flat $\Lambda$CDM cosmology with $H_{0} = 67.7$~km~ s$^{-1}$~Mpc$^{-1}$, $\Omega_{\Lambda}=0.714$ and $\Omega_{m} = 0.308$ \citep{2016A&A...594A..13P}, and a distance to the host of GW170817, NGC 4993, of 42.5 Mpc \citep{GW170817-MMAD}, i.e., $z\approx 0.0098$. For the computation of the upper bounds on gamma-ray flux and luminosity we assume a power law spectrum with a photon index of $-2$.  

We first performed a search for a transient counterpart within the 90\% contour of the final LIGO/Virgo \texttt{LALInference} map in the time window from \tgw + 1153 s to \tgw + 2027 s, the earliest interval in which the region was observed by the LAT, and no significant new sources were found. At the position of the optical counterpart~\citep{GCN21529, GW170817-MMAD} the value for the flux upper bound over this interval and in the 0.1--1 GeV energy range is 4.5$\times$10$^{-10}$ erg cm$^{-2}$ s$^{-1}$ (95\% confidence level), corresponding to an equivalent isotropic luminosity of approximately $9.7\times 10^{43}$ erg s$^{-1}$.

\begin{figure}
\begin{center}
\includegraphics[scale=0.5]{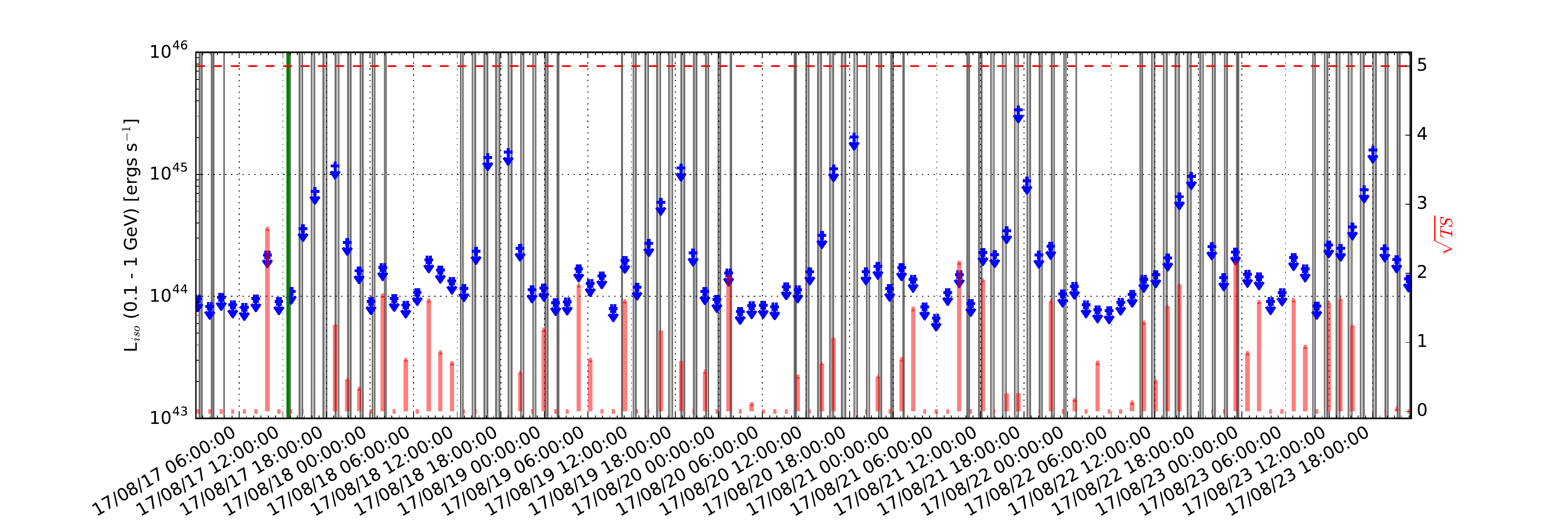}	
\includegraphics[scale=0.5]{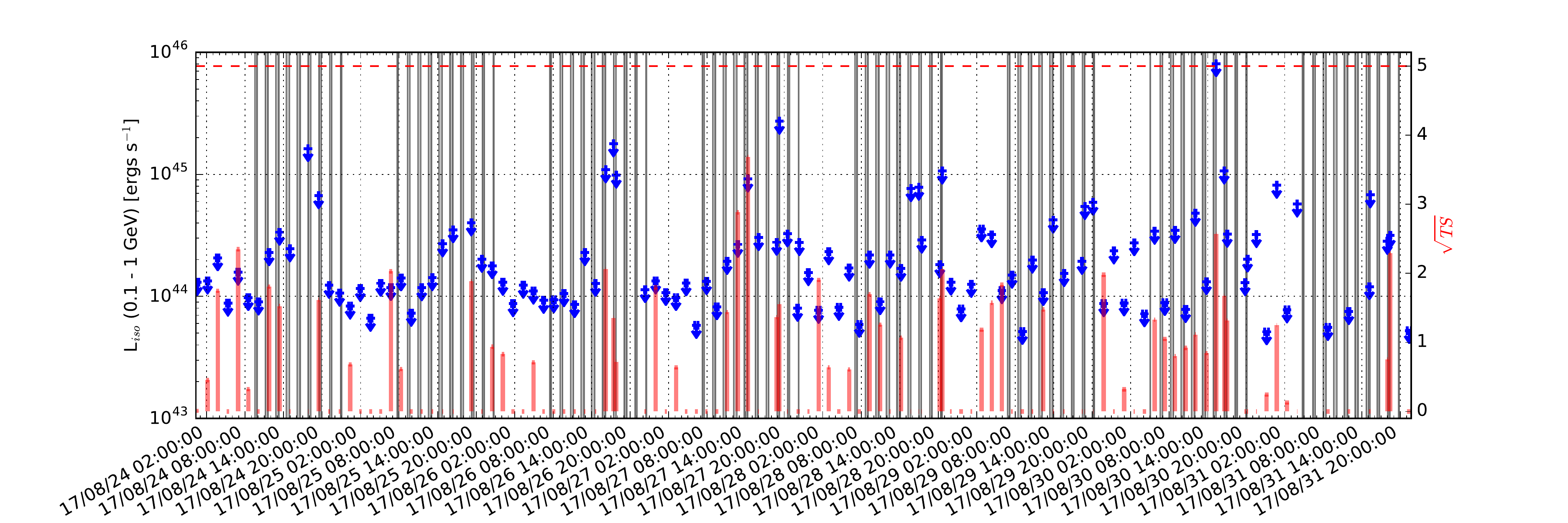}	
\end{center}
\caption{Upper bounds (blue points) for the first 2 weeks after \tem. The width of the blue points corresponds to the interval used in the analyses. The red bars of equivalent duration indicate the value of the significance (square root of the Test Statistic TS) in each interval, with the dashed red line representing a value corresponding to approximately 5$\sigma$ (TS=25).
The gray bands highlight the times when the LAT was in the SAA. The green vertical line in the upper panel is \tem.}
\label{fig:latetime}
\end{figure}

For typical GRBs where the viewing angle is smaller than the jet opening angle, the afterglow is coincident with the end of the prompt emission \citep{Berger2014,Avanzo:2015}, while for jets viewed off-axis, the onset of the afterglows is predicted to be of the order of few days up to 100 days \citep{Granot:2002, vanEerten2012}. As reported in \citet{GCN21787} and \citet{Troja:2017}, an X-ray source positionally coincident with the optical transient was detected at 8.9 days after the GW event, followed by a radio source detection \citet{GCN21814} suggesting the detection of an afterglow from a possible off-axis jet \citep{GW170817-MMAD}. 
Regardless of the origin of this emission, we monitored the source by performing a likelihood analysis in every interval of time after the trigger when the source was in the LAT field of view. The values of the flux upper bounds vary due to differences in exposures, as shown in Fig.~\ref{fig:latetime}. In a time period spanning from \tem-1 day to \tem+45 days these limits range between 9.7$\times$10$^{-11}$ to 3.7$\times$10$^{-8}$ erg cm$^{-2}$ s$^{-1}$ corresponding to a luminosity range of 2.1$\times$10$^{43}$ to 8.1$\times$10$^{45}$ erg s$^{-1}$ (0.1--1 GeV).

We also examined intervals spanning years before the trigger time. During normal operations, the LAT surveys the entire sky continuously and has observed the position of GRB\,170817A every $\sim 3$ hours since 2008, for a total of $\sim 55$ Ms of exposure. A search for a counterpart using data collected over the entire lifetime of the mission yielded no detection, returning an upper bound for the average 9-years flux of $1.32 \times 10^{-12}$ erg cm$^{-2}$ s$^{-1}$ (0.1--1 GeV), corresponding to a luminosity limit of $2.9 \times 10^{41}$ erg s$^{-1}$. We have also looked into the results of the diverse and complementary automatic techniques developed by the \Fermi-LAT team to continually search for transient events over a variety of timescales. For example, the \Fermi Automatic Science Processing (ASP) pipeline performs a search on six hour, one day, and one week timescales to identify candidate gamma-ray sources~\citep{2007AIPC..921..544C}. 
The candidate sources are subsequently reviewed through a more rigorous likelihood analysis by the Flare Advocate/Gamma-ray Sky Watcher (FA-GSW) pipeline, and positionally cross-checked with known cataloged gamma-ray objects. ASP allows for the detection of flux variations of known cataloged sources, as well as the detection of new unassociated gamma-ray transients. Similar to ASP, the \Fermi All-sky Variability Analysis (FAVA)~\citep{FAVA2013} searches for transients over 24 hour and one week timescales\footnote{https://fermi.gsfc.nasa.gov/ssc/data/access/lat/FAVA/}. FAVA is a blind photometric technique that looks for deviations from the expected flux in a grid of regions covering the entire sky. The observed long-term mission-averaged emission serves as reference for the expected flux, allowing the FAVA technique to be independent of any model of the gamma-ray sky. We searched through the full-mission dataset for transients detected by ASP and FAVA, positionally consistent with the optical counterpart. No sources with significance greater than 5$\sigma$ were found. The closest ASP transient is likely a blazar, 3FGL J1312.7-23, roughly $2^{\circ}$ from the optical counterpart and therefore unrelated.

\newpage

\section{GRB\,170817A in the context of other LAT detected SGRBs}

We can compare the properties of GRB\,170817A in the context of other GBM and LAT detected bursts. The GBM observations reported in \citet{Goldstein2017} show that the gamma-ray emission from GRB\,170817A was softer than that of typical SGRBs, with a peak in its $\nu F_{\nu}$ spectrum of $E_{\rm peak}=215 \pm 54$ keV, a value falling in the lowest $\sim 15^{th}$ percentile of the SGRB distribution. The burst fluence as measured in the 10--1000 keV energy range is dimmer than typical long bursts, but is consistent with those obtained for other GBM-detected SGRBs.

\begin{figure}[tb]
  \centering
  \includegraphics[width=0.5\textwidth]{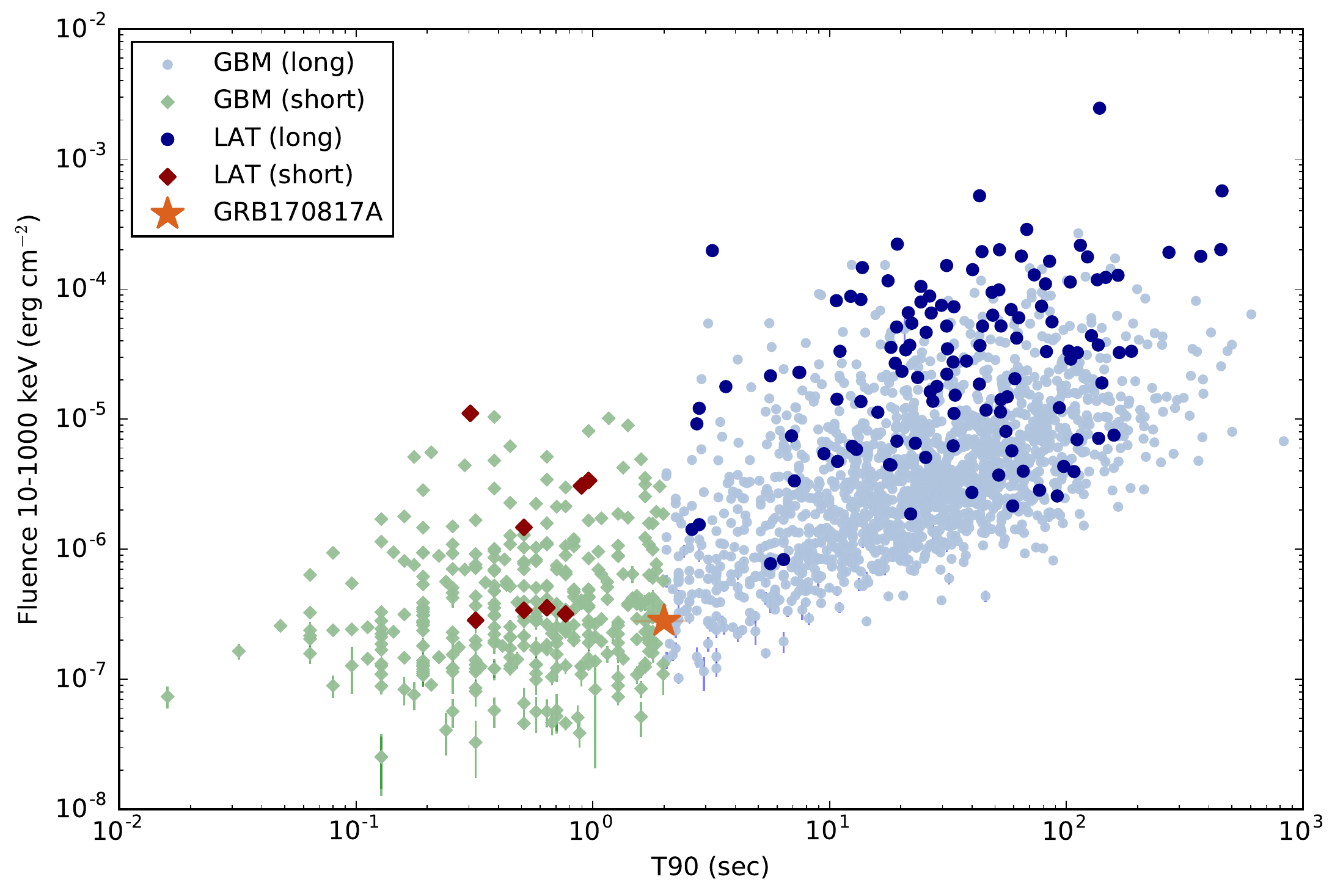}\hfill
  \includegraphics[width=0.5\textwidth]{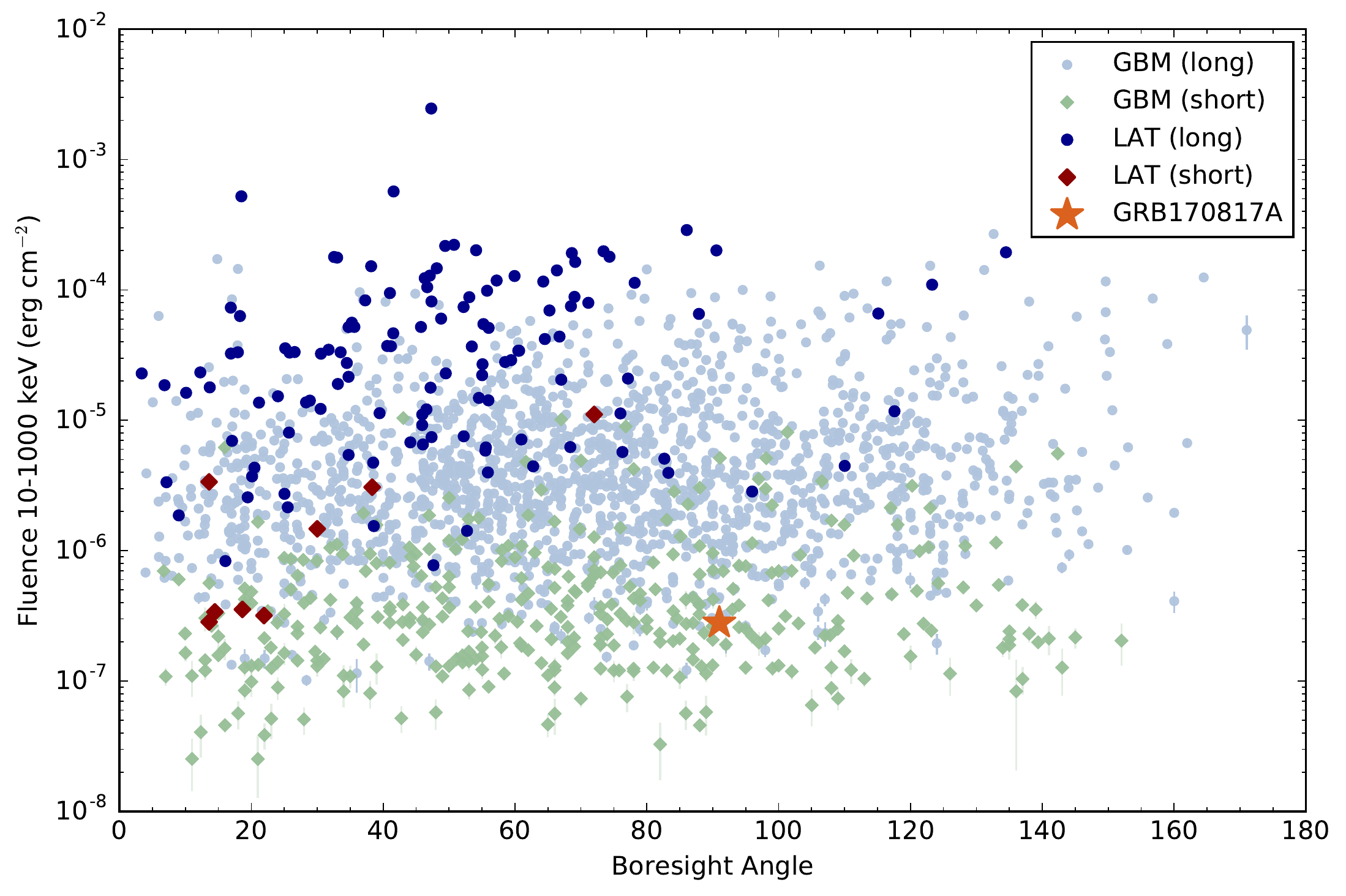}
  \caption{\emph{Left (Panel a):} The GBM energy fluence (10--1000 keV) plotted against the GBM duration for GRB~170817A (orange star) compared to the LAT-detected short (red diamonds) and long (blue circles) GRBs and non-detected short (light blue circles) and long (light green diamonds) populations of GRBs.\emph{Right (Panel b):} The GBM energy fluence (10--1000 keV) plotted against angle at which the burst occurred with respect to the LAT boresight at the time of trigger. The LAT detected GRBs with an off-axis angle $> 65^{\circ}$ were outside of the LAT field of view at the trigger time, but were detected after they entered the FOV at a later time.} 
  \label{fig:FluenceComparisons}
\end{figure}

The GBM has detected over 2000 GRBs in over 9 years of science operations, of which $\sim8\%$ have been detected at energies greater than 30 MeV by the LAT. Roughly $17\%$ of the GBM population is made up by SGRBs, with durations $T_{90} \leq 2$ s in the GBM energy range. The LAT detects $\sim5\%$ of the GBM-detected SGRBs.  The observed fluence in the 10--1000 keV energy range \citep{GBMBurstCatalog_6Years} and duration of the GBM and LAT detected populations can be seen in Figure~\ref{fig:FluenceComparisons}a. Fluences of SGRBs are generally lower than for long bursts, owing to their shorter durations. The LAT has detected 8 SGRBs with durations shorter than GRB\,170817A, four with comparably low fluence.

Figure~\ref{fig:FluenceComparisons}a shows the GBM fluence of the GBM and LAT detected populations plotted versus the angle at which the burst occurred with respect to the LAT boresight at the time of the GBM trigger (T0). Because the LAT sensitivity decreases significantly as a function of increasing boresight angle, the lowest fluence bursts are only detected by the LAT when they occur close to the instrument boresight. The location of the optical counterpart associated with GRB~170817A occurred $\sim$90$^{\circ}$ away from the LAT boresight at the time of the GBM trigger.  Four other LAT detected bursts (GRBs~160829, 140402, 090531, 081024) have fluence values as low as GRB~170817A, but each occurred within $\theta < 25^{\circ}$ of the LAT boresight, likely facilitated their detection.  
LAT bursts with boresight angles greater than $\theta > 65^{\circ}$ were either detected using the LAT Low Energy (LLE), which provides an additional sensitivity to off-axis photons, or were brought into the FOV through an automatic re-point request (ARR) of the spacecraft initiated by the GBM for high peak flux.

The LAT detects 1--2 SGRBs per year in the 100~MeV--100~GeV energy range \citep{2013ApJS..209...11A}. Their high-energy emission lasts longer than the prompt emission observed by the GBM, up to $\sim 200$ s after $T_0$ for the brightest cases. In order to compare GRB\,170817A with other LAT detected SGRBs we compute the flux upper bound in time interval \tgw + 1153 s to \tgw + 2027 s in the 100~MeV--100~GeV energy range, obtaining $2\times$10$^{-9}$~erg~cm$^{-2}$~s$^{-1}$ (corresponding to an isotropic luminosity of 4.3$\times10^{44}$~erg~s$^{-1}$). 
The flux upper bound value is above the expected flux at this time from an extrapolation of the power-law temporal decay of the brightest observed SGRB detected by the LAT to date, GRB\,090510~\citep{Ackermann2010, Razzaque2010}. Therefore, the lack of a detection of GRB\,170817A at $>1000$ s is consistent with previous non-detections at this time. 
We can compare the inferred luminosity upper bound of $4.3 \times 10^{44}$~erg~s$^{-1}$ for GRB~170817A to the luminosity of the extended emission for GRB~090510, also the only LAT-detected SGRB with known redshift (z=0.903). SGRBs in the LAT energy range are typically characterized by a power-law spectrum $\propto E^{-2}$ and a flux decaying as $\propto t^{-1}$ \citep{2013ApJS..209...11A}. By extrapolating the late-time light curve of GRB~090510~\citep{2013ApJS..209...11A}, we estimate the high-energy emission at $T_0$ + 1153~s in its source frame to be roughly $2\times10^{48}$~erg~s$^{-1}$. This effectively rules out late-time emission from GRB~170817A as luminous as GRB~090510.
The lack of such emission is not surprising, though, given that the prompt isotropic equivalent energy observed in the GBM for GRB~170817A of $E_{\rm iso} = 3.0 \pm 0.6 \times 10^{46}$~erg~\citep{Goldstein2017} is six orders of magnitude lower than the value estimated for GRB~090510. GRB~090510 radiated an isotropic equivalent energy of approximately $1.1~\times~10^{53}$~erg during its prompt emission observed by GBM and an additional $E_{\rm iso} = 5.5 \times~10^{52}$~erg in the high-energy extended emission detected by the LAT~\citep{2013ApJS..209...11A}. This indicates that the reason behind the under-luminous nature of GRB~170817A, e.g.\ off-axis emission, low Lorentz factor, intrinsically lower energy supply, etc.~\citep{jointpaper}, also affects the energetics of the component responsible for the high-energy emission observed by the LAT. If we assume a similar ratio between the prompt and extended $E_{\rm iso}$ for GRB~170817A as was determined for GRB~090510 ($\sim$50\%), then we would expect a luminosity for the extended emission for GRB~170817A at \tgw + 1153 s to be on the order of $5\times10^{41}$~erg~s$^{-1}$, well within our estimated luminosity upper bound.

\section{Prospects for future LAT detections and sensitivity study}

We now proceed to characterize the sensitivity of the LAT to SGRBs in general. This sensitivity is a function of the characteristics of the LAT, of its background, and of our search methodology. The background level and consequently the sensitivity varies across the sky, in particular as a function of Galactic latitude because of the bright diffuse emission associated with the plane of the Milky Way. We first present the computation of the sensitivity for the optical position of GRB~170817A, which corresponds to a middle Galactic latitude b=39$\fdg$296. We will then show how the sensitivity changes for typical positions in the sky at lower and higher Galactic latitude. As with all frequentist statistical tests, our method is characterized by the probability of false positives (Type I error, $\alpha$) and the probability of false negatives (Type II error, $\beta$). We consider an SGRB detected when it has a significance of at least 5$\sigma$ (TS$>$25), i.e., we fix $\alpha = 2.86 \times 10^{-7}$. Following \citet{2010ApJ...719..900K}, we can then compute the minimum flux that a point source must have in order to be detected above the 5$\sigma$ level with a given Type-II error probability. We fix for this computation $\beta = 0.5$.  We assume an exposure of 100~s, which is roughly the longest duration over which an SGRB has been detected by the LAT in the observer frame. We also assume, for simplicity, that the GRB is close to the axis of the LAT, where the effective area is maximized, and that the pointing is not changing with time. We use as background the Galactic and the Isotropic template provided by the LAT collaboration\footnote{Available at the \Fermi Science Support Center (https://fermi.gsfc.nasa.gov/ssc/)}, representing respectively the diffuse emission coming from the Milky Way and the isotropic component generated by particle misclassified as photons and by unresolved sources. We also include all known gamma-ray sources from the 3FGL catalog~\citep{3FGL} as part of the background. We then simulate repeatedly a point source with a spectrum $dN/dE \propto E^{-2}$, and vary the flux (averaged over the observation) until 50\% of the realizations of the simulated source are detected above 5$\sigma$. We ascertain that an average flux of $F_{s} = 9.5 \times 10^{-9}$~erg~cm$^{-2}$~s in the 0.1--100 GeV energy range is required to detect a source at mid-latitude with a type II error probability of $\beta$ = 50\% in a 100 s observation. The required flux increases (i.e., the sensitivity decreases) for a source on the Galactic plane by a factor of $\sim 2-3$ (depending on the Galactic longitude), and decreases (i.e. the sensitivity increases) by a factor of $\sim 2$ for a source at the Galactic poles, where the background is lower.

\begin{figure}[t]
\begin{center}
\includegraphics[scale=0.5]{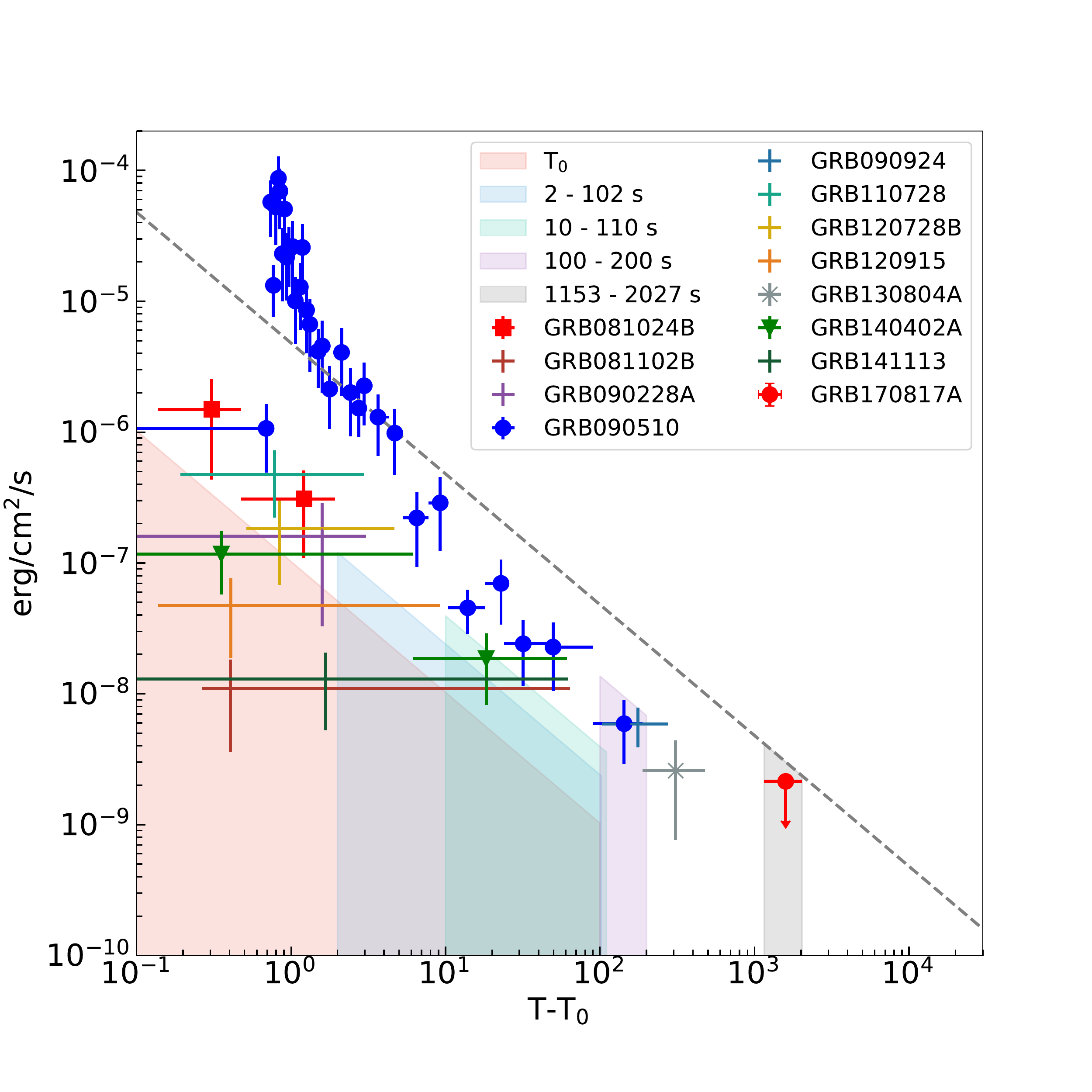}	
\end{center}
\caption{Light curves of the SGRBs detected by the LAT. We highlight  GRB~081024B (red squares), GRB~140402A (green triangles) GRB~130804 (gray cross) and GRB~090510 (blue circles). The fluence upper bound of GRB~170817A is also shown at the time of the first LAT observation (red circle). 
The shaded boxes represent the sensitivity to simulated sources detected with TS$>$25, 50\% of the time. The observation starts at  t$_{0}$, and 2, 10, 100$ֿ\;$s after the trigger and lasts 100$\;$s, as highlighted by the shaded regions. We also computed the sensitivity curve for an observation between 1153 and 2027$\;$s, as for GRB~170817A. This is also extrapolated back in time according to a $t^{-1}$ afterglow decay law (dashed gray line).}
\label{fig:shortGRBs}
\end{figure}

The starting time of a GRB observation is critical because the GRB flux fades rapidly, typically $\propto t^{-1}$. Therefore, we can detect fainter SGRBs the earlier we start observing. In Fig.~\ref{fig:shortGRBs} we show how the sensitivity of the LAT for sources at mid-Galactic latitudes changes for five different starting times. The first four shaded regions are for observations with a duration of 100~s, starting respectively at $T_{0}$, $T_{0} + 2$~s, $T_{0} + 10$~s, $T_{0} + 100$~s, while the last is between $T_{0} + 1153$~s and $T_{0} + 2027$~s after the trigger time as for GRB~170817A. For reference we also report the measurements for other SGRBs detected by the LAT, as well as the upper bound for GRB~170817A in the 100~MeV--100~GeV energy range. Among the sample of SGRBs detected by the LAT, we note that the fluences of GRB~081024B and GRB~140402A measured by the GBM in the 10 keV--1 MeV energy band are similar to the one of GRB~170817A. GRB~090510 is the brightest SGRB detected by the LAT so far and resulted in the detection of both its prompt and extended emission. The much dimmer GRB~130804A, on the other hand, was in the field of view at the time of the trigger, but was only detected by the LAT at $T_0 \sim 200$ s, constituting an example of delayed high-energy emission. The statistics are limited, but we can conclude that the LAT needs to start observing a source within 100--200 s to have a chance at detecting even the brightest of the LAT-detected SGRBs.

Because the LAT detection efficiency for SGRBs appears to decrease after 100--200 s, we next estimate the probability that the LAT will observe an SGRB within 100s of the trigger time during normal survey mode. The \Fermi-GBM observes $\sim$65\% of the sky, with the rest being occulted by the Earth. The LAT observed $\sim$35\% of SGRBs detected by the GBM within $\sim$100 s of the trigger. 
Hence, a continuation of this survey strategy indicates that on average, the LAT can observe (either a detection or upper bound) $\sim 23$\% of the full-sky SGRB population within 100 s from their GBM trigger. While the population of SGRBs is isotropically distributed, the population of SGRBs within the LIGO/Virgo horizon is likely not. Indeed, the expected horizon for LIGO/Virgo at design sensitivity to BH-NS and NS-NS mergers is estimated to be $\sim$190~Mpc~\citep{jointpaper}, and the distribution of galaxies that can potentially host an SGRB is not isotropic within such volume \citep{2MPZ}. However, we have performed simulations that showed that such anisotropy does not appreciably affect our observation rate.

The LAT detects $\sim$5\% of all GBM-detected SGRBs. If we assume the LAT will have the same efficiency for GRB/GW triggers and a rate of joint GBM/GW events of 1 (2) per year, we obtain at most a $\sim$5\% ($\sim$10\%) probability of detecting one or more GRB/GW with the LAT in one year, respectively. This assumes that GRB/GW events will be representative of the entire GBM-detected SGRB population when observed in gamma rays. Currently the \Fermi spacecraft autonomously slews to bring GRBs within the LAT field of view only when the GBM detects bursts of exceptionally high-peak flux. A modification of this strategy to repoint to lower-fluence SGRBs would provide increased exposure to dimmer events like GRB~170817A and increase the chances of detecting long-lived afterglow emission from such sources. Simulations have shown that this would allow the LAT to observe 35\% of SGRBs within 100~s, enhancing the probability of detecting one or more GRB/GW events per year to $\sim$7\% ($\sim$13\%) for a GBM/GW rate of 1 (2) per year. 

\section{Summary}

We present the \Fermi-LAT observations of the first confirmed LIGO/Virgo binary neutron star merger event GW170817 and the associated short gamma-ray burst GRB~170817A. Because the LAT was entering the SAA at the time of the LIGO/Virgo trigger, we can not place constraints on the high-energy (E $>$ 100 MeV) emission associated with the moment of binary coalescence. Instead we focus on constraining high-energy emission on longer timescales and report no candidate electromagnetic counterpart above $>100$ MeV on timescales of minutes, hours, and days after $t_{\rm GW}$. The resulting flux upper bound from the LAT is $4.5\times$10$^{-10}$ erg cm$^{-2}$ s$^{-1}$ in the 0.1--1 GeV range (and $2\times$10$^{-9}$ erg cm$^{-2}$ s$^{-1}$ in the 0.1--100 GeV energy range) covering the time interval \tgw + 1153 s to \tgw + 2027 s. This limit is above the expected flux at this time from an extrapolation of the power-law temporal decay of the brightest observed SGRB detected by the LAT to date, GRB~090510, which is also the only LAT SGRB with a measured redshift and much more distant ($z = 0.903$). The upper bound on the flux of GRB~170817A corresponds to a luminosity upper bound in the 0.1 -- 1 GeV energy range of 9.7$\times10^{43}$ erg s$^{-1}$ (or 4.3$\times10^{44}$ erg s$^{-1}$ in the 0.1--100 GeV energy range), significantly lower than the luminosity for GRB~090510 obtained by extrapolating its light curve to $1153$ s after the trigger time ($4\times10^{49}$ erg s$^{-1}$). This effectively rules out emission for GRB~170817A as luminous as GRB~090510. However, this is not surprising as the total energy output ($E_{\rm iso}$) during the prompt emission, as measured by the GBM , was 6 orders of magnitude lower for GRB~170817A than for GRB~090510. This might indicate an intrinsically less energetic event or a jet observed off-axis instead of on-axis, or both.

The host galaxy of GRB~170817A exhibits some features of an Active Galactic Nuclei \citep{GW170817-MMAD}. We have therefore also looked for a possible high-energy signal in the $\sim 55$ Ms of exposure that the LAT has accumulated at the position of GRB~170817A over the life of the \Fermi mission. We do not detect any signal from the host galaxy over the entire mission, placing a luminosity upper bound of $2.9 \times 10^{41}$ erg s$^{-1}$ (0.1 -- 100 GeV).  We also did not detect any signal on time scales of minutes, hours or days in the $\sim 9$ years before the trigger time.

The LAT has detected SGRBs with GBM fluences that are comparable to GRB~170817A. Therefore, it is possible for the LAT to detect GRBs characterized by a low-energy prompt emission as faint as GRB~170817A, provided that they have a ratio between low-energy prompt emission and high-energy extended emission similar to previously detected bursts and are observed relatively close to the boresight of the instrument. We have estimated the sensitivity of the LAT to SGRBs and compared it to these previous observations, concluding that the LAT would need to begin observations within $\sim$ 100 s from the trigger time. 

Finally, under the current observation strategy the LAT should observe $\sim 23$\% of the population of SGRBs within 100 s from the trigger, providing meaningful constraints. At the current LAT detection efficiency of $\sim$5\% of all GBM-detected SGRBs~\citep{2013ApJS..209...11A}, we estimate that a joint GRB/GW rate of 1 (2) per year would result in at most a $\sim$5\% ($\sim$10\%) probability of LAT detecting one or more GRB/GW in one year. This assumes that GRB/GW events are similar to the previously detected population of SGRBs when observed in gamma rays. The percentage of SGRBs observed by the LAT within 100 s could increase to at least $\sim 35$\% by automatically slewing the spacecraft to bring all GBM-detected SGRBs to the center of the LAT FOV. This would increase the probability of detecting one or more GRB/GW events per year to 7\% (13\%). These results could be compared with \citet{2016JCAP...11..056P}.

The detection of gravitational waves and an associated EM signal from the coalescence of compact binary system has initiated a new phase in high-energy astrophysics. 
The discovery of GW170817/GRB~170817A strongly supports the conjectured association of SGRBs with merging neutron stars. Multi-messenger study of future events can verify this  and should impact our understanding of many astrophysical processes - general relativistic dynamics, the equation of state of cold nuclear matter, relativistic jet formation by compact objects, particle acceleration, radioactive heating of the expanding debris (as in kilonovae) and cosmography. We already know that the observations depend strongly upon orientation, which can now be determined, at least in principle. Other factors, in particular the initial neutron star masses, spins and magnetization and the environment may also play a role but ought not to be very important. Presuming that there will be an improvement in LIGO-Virgo sensitivity which results in a much larger detection rate and given what we already know from GBM and LAT, there is every reason to believe that gamma-ray observations will drive future developments in this exciting field.

\bigskip
\centerline{We dedicate this paper to our late friend and colleague Neil Gehrels,} \centerline{whose curiosity and enthusiasm inspired so much of our work.}
\medskip
\noindent The \textit{Fermi} LAT Collaboration acknowledges generous ongoing support
from a number of agencies and institutes that have supported both the
development and the operation of the LAT as well as scientific data analysis.
These include the National Aeronautics and Space Administration and the
Department of Energy in the United States, the Commissariat \`a l'Energie Atomique
and the Centre National de la Recherche Scientifique / Institut National de Physique
Nucl\'eaire et de Physique des Particules in France, the Agenzia Spaziale Italiana
and the Istituto Nazionale di Fisica Nucleare in Italy, the Ministry of Education,
Culture, Sports, Science and Technology (MEXT), High Energy Accelerator Research
Organization (KEK) and Japan Aerospace Exploration Agency (JAXA) in Japan, and
the K.~A.~Wallenberg Foundation, the Swedish Research Council and the
Swedish National Space Board in Sweden.
 
Additional support for science analysis during the operations phase is gratefully
acknowledged from the Istituto Nazionale di Astrofisica in Italy and the Centre
National d'\'Etudes Spatiales in France. This work performed in part under DOE
Contract DE-AC02-76SF00515.

\bibliographystyle{aasjournal}
\bibliography{bibliography.bib}

\begin{thebibliography}{}
\expandafter\ifx\csname natexlab\endcsname\relax\def\natexlab#1{#1}\fi
\providecommand{\url}[1]{\href{#1}{#1}}

\bibitem[{Abbott {et~al.}(2016)Abbott, Abbott, Abbott, Abernathy, Acernese,
  Ackley, Adams, Adams, Addesso, Adhikari, Adya, Affeldt, Agathos, Agatsuma,
  Aggarwal, Aguiar, Aiello, Ain, Ajith, Allen, Allocca, Altin, Anderson,
  Anderson, Arai, Araya, Arceneaux, Areeda, Arnaud, Arun, Ascenzi, Ashton, Ast,
  Aston, Astone, Aufmuth, Aulbert, Babak, Bacon, Bader, Baker, Baldaccini,
  Ballardin, Ballmer, Barayoga, Barclay, Barish, Barker, Barone, Barr,
  Barsotti, Barsuglia, Barta, Bartlett, Bartos, Bassiri, Basti, Batch, Baune,
  Bavigadda, Bazzan, Behnke, Bejger, Bell, Bell, Berger, Bergman, Bergmann,
  Berry, Bersanetti, Bertolini, Betzwieser, Bhagwat, Bhandare, Bilenko,
  Billingsley, Birch, Birney, Biscans, Bisht, Bitossi, Biwer, Bizouard,
  Blackburn, Blair, Blair, Blair, Bloemen, Bock, Bodiya, Boer, Bogaert, Bogan,
  Bohe, Bojtos, Bond, Bondu, Bonnand, Boom, Bork, Boschi, Bose, Bouffanais,
  Bozzi, Bradaschia, Brady, Braginsky, Branchesi, Brau, Briant, Brillet,
  Brinkmann, Brisson, Brockill, Brooks, Brown, Brown, Brown, Buchanan, Buikema,
  Bulik, Bulten, Buonanno, Buskulic, Buy, Byer, Cadonati, Cagnoli, Cahillane,
  Calder\'on~Bustillo, Callister, Calloni, Camp, Cannon, Cao, Capano, Capocasa,
  Carbognani, Caride, Casanueva~Diaz, Casentini, Caudill, Cavagli\`a, Cavalier,
  Cavalieri, Cella, Cepeda, Cerboni~Baiardi, Cerretani, Cesarini, Chakraborty,
  Chalermsongsak, Chamberlin, Chan, Chao, Charlton, Chassande-Mottin, Chen,
  Chen, Cheng, Chincarini, Chiummo, Cho, Cho, Chow, Christensen, Chu, Chua,
  Chung, Ciani, Clara, Clark, Cleva, Coccia, Cohadon, Colla, Collette,
  Cominsky, Constancio, Conte, Conti, Cook, Corbitt, Cornish, Corsi, Cortese,
  Costa, Coughlin, Coughlin, Coulon, Countryman, Couvares, Cowan, Coward,
  Cowart, Coyne, Coyne, Craig, Creighton, Cripe, Crowder, Cumming, Cunningham,
  Cuoco, Dal~Canton, Danilishin, D'Antonio, Danzmann, Darman, Dattilo, Dave,
  Daveloza, Davier, Davies, Daw, Day, DeBra, Debreczeni, Degallaix,
  De~Laurentis, Del\'eglise, Del~Pozzo, Denker, Dent, Dereli, Dergachev,
  DeRosa, De~Rosa, DeSalvo, Dhurandhar, D\'{\i}az, Di~Fiore, Di~Giovanni,
  Di~Lieto, Di~Pace, Di~Palma, Di~Virgilio, Dojcinoski, Dolique, Donovan,
  Dooley, Doravari, Douglas, Downes, Drago, Drever, Driggers, Du, Ducrot,
  Dwyer, Edo, Edwards, Effler, Eggenstein, Ehrens, Eichholz, Eikenberry,
  Engels, Essick, Etzel, Evans, Evans, Everett, Factourovich, Fafone, Fair,
  Fairhurst, Fan, Fang, Farinon, Farr, Farr, Favata, Fays, Fehrmann, Fejer,
  Ferrante, Ferreira, Ferrini, Fidecaro, Fiori, Fiorucci, Fisher, Flaminio,
  Fletcher, Fournier, Franco, Frasca, Frasconi, Frei, Freise, Frey, Frey,
  Fricke, Fritschel, Frolov, Fulda, Fyffe, Gabbard, Gair, Gammaitoni, Gaonkar,
  Garufi, Gatto, Gaur, Gehrels, Gemme, Gendre, Genin, Gennai, George, Gergely,
  Germain, Ghosh, Ghosh, Giaime, Giardina, Giazotto, Gill, Glaefke, Goetz,
  Goetz, Gondan, Gonz\'alez, Gonzalez~Castro, Gopakumar, Gordon, Gorodetsky,
  Gossan, Gosselin, Gouaty, Graef, Graff, Granata, Grant, Gras, Gray, Greco,
  Green, Groot, Grote, Grunewald, Guidi, Guo, Gupta, Gupta, Gushwa, Gustafson,
  Gustafson, Hacker, Hall, Hall, Hammond, Haney, Hanke, Hanks, Hanna, Hannam,
  Hanson, Hardwick, Haris, Harms, Harry, Harry, Hart, Hartman, Haster,
  Haughian, Heidmann, Heintze, Heitmann, Hello, Hemming, Hendry, Heng, Hennig,
  Heptonstall, Heurs, Hild, Hoak, Hodge, Hofman, Hollitt, Holt, Holz, Hopkins,
  Hosken, Hough, Houston, Howell, Hu, Huang, Huerta, Huet, Hughey, Husa,
  Huttner, Huynh-Dinh, Idrisy, Indik, Ingram, Inta, Isa, Isac, Isi, Islas,
  Isogai, Iyer, Izumi, Jacqmin, Jang, Jani, Jaranowski, Jawahar,
  Jim\'enez-Forteza, Johnson, Jones, Jones, Jonker, Ju, Kalaghatgi, Kalogera,
  Kandhasamy, Kang, Kanner, Karki, Kasprzack, Katsavounidis, Katzman, Kaufer,
  Kaur, Kawabe, Kawazoe, K\'ef\'elian, Kehl, Keitel, Kelley, Kells, Kennedy,
  Key, Khalaidovski, Khalili, Khan, Khan, Khan, Khazanov, Kijbunchoo, Kim, Kim,
  Kim, Kim, Kim, Kim, King, King, Kinzel, Kissel, Kleybolte, Klimenko,
  Koehlenbeck, Kokeyama, Koley, Kondrashov, Kontos, Korobko, Korth, Kowalska,
  Kozak, Kringel, Kr\'olak, Krueger, Kuehn, Kumar, Kuo, Kutynia, Lackey,
  Landry, Lange, Lantz, Lasky, Lazzarini, Lazzaro, Leaci, Leavey, Lebigot, Lee,
  Lee, Lee, Lee, Lenon, Leonardi, Leong, Leroy, Letendre, Levin, Levine, Li,
  Libson, Littenberg, Lockerbie, Logue, Lombardi, Lord, Lorenzini, Loriette,
  Lormand, Losurdo, Lough, L\"uck, Lundgren, Luo, Lynch, Ma, MacDonald,
  Machenschalk, MacInnis, Macleod, Maga\~na Sandoval, Magee, Mageswaran,
  Majorana, Maksimovic, Malvezzi, Man, Mandel, Mandic, Mangano, Mansell,
  Manske, Mantovani, Marchesoni, Marion, M\'arka, M\'arka, Markosyan, Maros,
  Martelli, Martellini, Martin, Martin, Martynov, Marx, Mason, Masserot,
  Massinger, Masso-Reid, Matichard, Matone, Mavalvala, Mazumder, Mazzolo,
  McCarthy, McClelland, McCormick, McGuire, McIntyre, McIver, McManus,
  McWilliams, Meacher, Meadors, Meidam, Melatos, Mendell, Mendoza-Gandara,
  Mercer, Merilh, Merzougui, Meshkov, Messenger, Messick, Meyers, Mezzani,
  Miao, Michel, Middleton, Mikhailov, Milano, Miller, Millhouse, Minenkov,
  Ming, Mirshekari, Mishra, Mitra, Mitrofanov, Mitselmakher, Mittleman, Moggi,
  Mohan, Mohapatra, Montani, Moore, Moore, Moraru, Moreno, Morriss, Mossavi,
  Mours, Mow-Lowry, Mueller, Mueller, Muir, Mukherjee, Mukherjee, Mukherjee,
  Mukund, Mullavey, Munch, Murphy, Murray, Mytidis, Nardecchia, Naticchioni,
  Nayak, Necula, Nedkova, Nelemans, Neri, Neunzert, Newton, Nguyen, Nielsen,
  Nissanke, Nitz, Nocera, Nolting, Normandin, Nuttall, Oberling, Ochsner,
  O'Dell, Oelker, Ogin, Oh, Oh, Ohme, Oliver, Oppermann, Oram, O'Reilly,
  O'Shaughnessy, Ottaway, Ottens, Overmier, Owen, Pai, Pai, Palamos, Palashov,
  Palomba, Pal-Singh, Pan, Pankow, Pannarale, Pant, Paoletti, Paoli, Papa,
  Paris, Parker, Pascucci, Pasqualetti, Passaquieti, Passuello, Patricelli,
  Patrick, Pearlstone, Pedraza, Pedurand, Pekowsky, Pele, Penn, Perreca,
  Phelps, Piccinni, Pichot, Piergiovanni, Pierro, Pillant, Pinard, Pinto,
  Pitkin, Poggiani, Popolizio, Post, Powell, Prasad, Predoi, Premachandra,
  Prestegard, Price, Prijatelj, Principe, Privitera, Prodi, Prokhorov, Puncken,
  Punturo, Puppo, P\"urrer, Qi, Qin, Quetschke, Quintero, Quitzow-James, Raab,
  Rabeling, Radkins, Raffai, Raja, Rakhmanov, Rapagnani, Raymond, Razzano, Re,
  Read, Reed, Regimbau, Rei, Reid, Reitze, Rew, Reyes, Ricci, Riles, Robertson,
  Robie, Robinet, Rocchi, Rolland, Rollins, Roma, Romano, Romanov, Romie,
  Rosi\ifmmode~\acute{n}\else \'{n}\fi{}ska, Rowan, R\"udiger, Ruggi, Ryan,
  Sachdev, Sadecki, Sadeghian, Salconi, Saleem, Salemi, Samajdar, Sammut,
  Sanchez, Sandberg, Sandeen, Sanders, Sassolas, Sathyaprakash, Saulson,
  Sauter, Savage, Sawadsky, Schale, Schilling, Schmidt, Schmidt, Schnabel,
  Schofield, Sch\"onbeck, Schreiber, Schuette, Schutz, Scott, Scott, Sellers,
  Sengupta, Sentenac, Sequino, Sergeev, Serna, Setyawati, Sevigny, Shaddock,
  Shah, Shahriar, Shaltev, Shao, Shapiro, Shawhan, Sheperd, Shoemaker,
  Shoemaker, Siellez, Siemens, Sigg, Silva, Simakov, Singer, Singer, Singh,
  Singh, Singhal, Sintes, Slagmolen, Smith, Smith, Smith, Son, Sorazu,
  Sorrentino, Souradeep, Srivastava, Staley, Steinke, Steinlechner,
  Steinlechner, Steinmeyer, Stephens, Stone, Strain, Straniero, Stratta,
  Strauss, Strigin, Sturani, Stuver, Summerscales, Sun, Sutton, Swinkels,
  Szczepa\ifmmode~\acute{n}\else \'{n}\fi{}czyk, Tacca, Talukder, Tanner,
  T\'apai, Tarabrin, Taracchini, Taylor, Theeg, Thirugnanasambandam, Thomas,
  Thomas, Thomas, Thorne, Thorne, Thrane, Tiwari, Tiwari, Tokmakov, Tomlinson,
  Tonelli, Torres, Torrie, T\"oyr\"a, Travasso, Traylor, Trifir\`o, Tringali,
  Trozzo, Tse, Turconi, Tuyenbayev, Ugolini, Unnikrishnan, Urban, Usman,
  Vahlbruch, Vajente, Valdes, van Bakel, van Beuzekom, van~den Brand, Van
  Den~Broeck, Vander-Hyde, van~der Schaaf, van Heijningen, van Veggel, Vardaro,
  Vass, Vas\'uth, Vaulin, Vecchio, Vedovato, Veitch, Veitch, Venkateswara,
  Verkindt, Vetrano, Vicer\'e, Vinciguerra, Vine, Vinet, Vitale, Vo, Vocca,
  Vorvick, Voss, Vousden, Vyatchanin, Wade, Wade, Wade, Walker, Wallace, Walsh,
  Wang, Wang, Wang, Wang, Wang, Ward, Warner, Was, Weaver, Wei, Weinert,
  Weinstein, Weiss, Welborn, Wen, We\ss{}els, Westphal, Wette, Whelan,
  Whitcomb, White, Whiting, Williams, Williamson, Willis, Willke, Wimmer,
  Winkler, Wipf, Wittel, Woan, Worden, Wright, Wu, Yablon, Yam, Yamamoto,
  Yancey, Yap, Yu, Yvert, Zadro\ifmmode~\dot{z}\else \.{z}\fi{}ny, Zangrando,
  Zanolin, Zendri, Zevin, Zhang, Zhang, Zhang, Zhang, Zhao, Zhou, Zhou, Zhu,
  Zucker, Zuraw, \& Zweizig}]{AdvLIGO}
Abbott, B.~P., Abbott, R., Abbott, T.~D., {et~al.} 2016, Phys. Rev. Lett., 116,
  131103

\bibitem[{Abbott {et~al.}(2017)}]{GW170817}
Abbott, B.~P., {et~al.} 2017, \prl, 119, 161101

\bibitem[{{Abbott} {et~al.}(2017{\natexlab{a}})}]{GW170817-MMAD}
{Abbott}, B.~P., {et~al.} 2017{\natexlab{a}}, \apjl, in press.,
  doi:10.3847/2041-8213/aa91c9

\bibitem[{{Abbott} {et~al.}(2017{\natexlab{b}})}]{jointpaper}
{Abbott}, B.~P., G., {et~al.} 2017{\natexlab{b}}, \apjl, 828,
  doi:10.3847/2041-8213/

\bibitem[{{Acero} {et~al.}(2015){Acero}, {Ackermann}, {Ajello}, {Albert},
  {Atwood}, {Axelsson}, {Baldini}, {Ballet}, {Barbiellini}, {Bastieri},
  {Belfiore}, {Bellazzini}, {Bissaldi}, {Blandford}, {Bloom}, {Bogart},
  {Bonino}, {Bottacini}, {Bregeon}, {Britto}, {Bruel}, {Buehler}, {Burnett},
  {Buson}, {Caliandro}, {Cameron}, {Caputo}, {Caragiulo}, {Caraveo},
  {Casandjian}, {Cavazzuti}, {Charles}, {Chaves}, {Chekhtman}, {Cheung},
  {Chiang}, {Chiaro}, {Ciprini}, {Claus}, {Cohen-Tanugi}, {Cominsky}, {Conrad},
  {Cutini}, {D'Ammando}, {de Angelis}, {DeKlotz}, {de Palma}, {Desiante},
  {Digel}, {Di Venere}, {Drell}, {Dubois}, {Dumora}, {Favuzzi}, {Fegan},
  {Ferrara}, {Finke}, {Franckowiak}, {Fukazawa}, {Funk}, {Fusco}, {Gargano},
  {Gasparrini}, {Giebels}, {Giglietto}, {Giommi}, {Giordano}, {Giroletti},
  {Glanzman}, {Godfrey}, {Grenier}, {Grondin}, {Grove}, {Guillemot}, {Guiriec},
  {Hadasch}, {Harding}, {Hays}, {Hewitt}, {Hill}, {Horan}, {Iafrate}, {Jogler},
  {J{\'o}hannesson}, {Johnson}, {Johnson}, {Johnson}, {Johnson}, {Kamae},
  {Kataoka}, {Katsuta}, {Kuss}, {La Mura}, {Landriu}, {Larsson}, {Latronico},
  {Lemoine-Goumard}, {Li}, {Li}, {Longo}, {Loparco}, {Lott}, {Lovellette},
  {Lubrano}, {Madejski}, {Massaro}, {Mayer}, {Mazziotta}, {McEnery},
  {Michelson}, {Mirabal}, {Mizuno}, {Moiseev}, {Mongelli}, {Monzani},
  {Morselli}, {Moskalenko}, {Murgia}, {Nuss}, {Ohno}, {Ohsugi}, {Omodei},
  {Orienti}, {Orlando}, {Ormes}, {Paneque}, {Panetta}, {Perkins},
  {Pesce-Rollins}, {Piron}, {Pivato}, {Porter}, {Racusin}, {Rando}, {Razzano},
  {Razzaque}, {Reimer}, {Reimer}, {Reposeur}, {Rochester}, {Romani},
  {Salvetti}, {S{\'a}nchez-Conde}, {Saz Parkinson}, {Schulz}, {Siskind},
  {Smith}, {Spada}, {Spandre}, {Spinelli}, {Stephens}, {Strong}, {Suson},
  {Takahashi}, {Takahashi}, {Tanaka}, {Thayer}, {Thayer}, {Thompson},
  {Tibaldo}, {Tibolla}, {Torres}, {Torresi}, {Tosti}, {Troja}, {Van Klaveren},
  {Vianello}, {Winer}, {Wood}, {Wood}, {Zimmer}, \& {Fermi-LAT
  Collaboration}}]{3FGL}
{Acero}, F., {Ackermann}, M., {Ajello}, M., {et~al.} 2015, \apjs, 218, 23

\bibitem[{{Ackermann} {et~al.}(2010){Ackermann}, {Asano}, {Atwood}, {Axelsson},
  {Baldini}, {Ballet}, {Barbiellini}, {Baring}, {Bastieri}, {Bechtol},
  {Bellazzini}, {Berenji}, {Bhat}, {Bissaldi}, {Blandford}, {Bloom},
  {Bonamente}, {Borgland}, {Bouvier}, {Bregeon}, {Brez}, {Briggs}, {Brigida},
  {Bruel}, {Buson}, {Caliandro}, {Cameron}, {Caraveo}, {Carrigan},
  {Casandjian}, {Cecchi}, {{\c C}elik}, {Charles}, {Chiang}, {Ciprini},
  {Claus}, {Cohen-Tanugi}, {Connaughton}, {Conrad}, {Dermer}, {de Palma},
  {Dingus}, {Silva}, {Drell}, {Dubois}, {Dumora}, {Farnier}, {Favuzzi},
  {Fegan}, {Finke}, {Focke}, {Frailis}, {Fukazawa}, {Fusco}, {Gargano},
  {Gasparrini}, {Gehrels}, {Germani}, {Giglietto}, {Giordano}, {Glanzman},
  {Godfrey}, {Granot}, {Grenier}, {Grondin}, {Grove}, {Guiriec}, {Hadasch},
  {Harding}, {Hays}, {Horan}, {Hughes}, {J{\'o}hannesson}, {Johnson}, {Kamae},
  {Katagiri}, {Kataoka}, {Kawai}, {Kippen}, {Kn{\"o}dlseder}, {Kocevski},
  {Kouveliotou}, {Kuss}, {Lande}, {Latronico}, {Lemoine-Goumard}, {Llena
  Garde}, {Longo}, {Loparco}, {Lott}, {Lovellette}, {Lubrano}, {Makeev},
  {Mazziotta}, {McEnery}, {McGlynn}, {Meegan}, {M{\'e}sz{\'a}ros}, {Michelson},
  {Mitthumsiri}, {Mizuno}, {Moiseev}, {Monte}, {Monzani}, {Moretti},
  {Morselli}, {Moskalenko}, {Murgia}, {Nakajima}, {Nakamori}, {Nolan},
  {Norris}, {Nuss}, {Ohno}, {Ohsugi}, {Omodei}, {Orlando}, {Ormes}, {Ozaki},
  {Paciesas}, {Paneque}, {Panetta}, {Parent}, {Pelassa}, {Pepe},
  {Pesce-Rollins}, {Piron}, {Preece}, {Rain{\`o}}, {Rando}, {Razzano},
  {Razzaque}, {Reimer}, {Ritz}, {Rodriguez}, {Roth}, {Ryde}, {Sadrozinski},
  {Sander}, {Scargle}, {Schalk}, {Sgr{\`o}}, {Siskind}, {Smith}, {Spandre},
  {Spinelli}, {Stamatikos}, {Stecker}, {Strickman}, {Suson}, {Tajima},
  {Takahashi}, {Takahashi}, {Tanaka}, {Thayer}, {Thayer}, {Thompson},
  {Tibaldo}, {Toma}, {Torres}, {Tosti}, {Tramacere}, {Uchiyama}, {Uehara},
  {Usher}, {van der Horst}, {Vasileiou}, {Vilchez}, {Vitale}, {von Kienlin},
  {Waite}, {Wang}, {Wilson-Hodge}, {Winer}, {Wu}, {Yamazaki}, {Yang}, {Ylinen},
  \& {Ziegler}}]{Ackermann2010}
{Ackermann}, M., {Asano}, K., {Atwood}, W.~B., {et~al.} 2010, \apj, 716, 1178

\bibitem[{{Ackermann} {et~al.}(2013{\natexlab{a}}){Ackermann}, {Ajello},
  {Asano}, {Baldini}, {Barbiellini}, {Baring}, {Bastieri}, {Bellazzini},
  {Blandford}, {Bonamente}, {Borgland}, {Bottacini}, {Bregeon}, {Brigida},
  {Bruel}, {Buehler}, {Buson}, {Caliandro}, {Cameron}, {Caraveo}, {Cecchi},
  {Charles}, {Chaves}, {Chekhtman}, {Chiang}, {Ciprini}, {Claus},
  {Cohen-Tanugi}, {Conrad}, {Cutini}, {D'Ammando}, {de Angelis}, {de Palma},
  {Dermer}, {Silva}, {Drell}, {Drlica-Wagner}, {Favuzzi}, {Fegan}, {Focke},
  {Franckowiak}, {Fukazawa}, {Fusco}, {Gargano}, {Gasparrini}, {Gehrels},
  {Giglietto}, {Giordano}, {Giroletti}, {Glanzman}, {Godfrey}, {Granot},
  {Greiner}, {Grenier}, {Grove}, {Guiriec}, {Hadasch}, {Hanabata}, {Hayashida},
  {Hays}, {Hughes}, {Jackson}, {Jogler}, {J{\'o}hannesson}, {Johnson},
  {Kn{\"o}dlseder}, {Kocevski}, {Kuss}, {Lande}, {Larsson}, {Latronico},
  {Longo}, {Loparco}, {Lovellette}, {Lubrano}, {Mazziotta}, {McEnery},
  {Mehault}, {M{\'e}sz{\'a}ros}, {Michelson}, {Mitthumsiri}, {Mizuno}, {Monte},
  {Monzani}, {Moretti}, {Morselli}, {Moskalenko}, {Murgia}, {Naumann-Godo},
  {Norris}, {Nuss}, {Nymark}, {Ohno}, {Ohsugi}, {Omodei}, {Orienti}, {Orlando},
  {Paneque}, {Perkins}, {Pesce-Rollins}, {Piron}, {Pivato}, {Racusin},
  {Rain{\`o}}, {Rando}, {Razzano}, {Razzaque}, {Reimer}, {Reimer}, {Romoli},
  {Roth}, {Ryde}, {Sanchez}, {Sgr{\`o}}, {Siskind}, {Sonbas}, {Spinelli},
  {Stamatikos}, {Takahashi}, {Tanaka}, {Thayer}, {Thayer}, {Tibaldo},
  {Tinivella}, {Tosti}, {Troja}, {Usher}, {Vandenbroucke}, {Vasileiou},
  {Vianello}, {Vitale}, {Waite}, {Winer}, {Wood}, {Yang}, {Gruber}, {Bhat},
  {Bissaldi}, {Briggs}, {Burgess}, {Connaughton}, {Foley}, {Kippen},
  {Kouveliotou}, {McBreen}, {McGlynn}, {Paciesas}, {Pelassa}, {Preece}, {Rau},
  {van der Horst}, {von Kienlin}, {Kann}, {Filgas}, {Klose}, {Kr{\"u}hler},
  {Fukui}, {Sako}, {Tristram}, {Oates}, {Ukwatta}, \&
  {Littlejohns}}]{Ackermann2013}
{Ackermann}, M., {Ajello}, M., {Asano}, K., {et~al.} 2013{\natexlab{a}}, \apj,
  763, 71

\bibitem[{{Ackermann} {et~al.}(2013{\natexlab{b}}){Ackermann}, {Ajello},
  {Albert}, {Allafort}, {Antolini}, {Baldini}, {Ballet}, {Barbiellini},
  {Bastieri}, {Bechtol}, {Bellazzini}, {Blandford}, {Bloom}, {Bonamente},
  {Bottacini}, {Bouvier}, {Brandt}, {Bregeon}, {Brigida}, {Bruel}, {Buehler},
  {Buson}, {Caliandro}, {Cameron}, {Caraveo}, {Cavazzuti}, {Cecchi}, {Charles},
  {Chekhtman}, {Cheung}, {Chiang}, {Chiaro}, {Ciprini}, {Claus},
  {Cohen-Tanugi}, {Conrad}, {Cutini}, {Dalton}, {D'Ammando}, {de Angelis}, {de
  Palma}, {Dermer}, {Di Venere}, {Drell}, {Drlica-Wagner}, {Favuzzi}, {Fegan},
  {Ferrara}, {Focke}, {Franckowiak}, {Fukazawa}, {Funk}, {Fusco}, {Gargano},
  {Gasparrini}, {Germani}, {Giglietto}, {Giordano}, {Giroletti}, {Glanzman},
  {Godfrey}, {Grenier}, {Grondin}, {Grove}, {Guiriec}, {Hadasch}, {Hanabata},
  {Harding}, {Hayashida}, {Hays}, {Hewitt}, {Hill}, {Horan}, {Hou}, {Hughes},
  {Inoue}, {Jackson}, {Jogler}, {J{\'o}hannesson}, {Johnson}, {Kamae},
  {Kataoka}, {Kawano}, {Kn{\"o}dlseder}, {Kuss}, {Lande}, {Larsson},
  {Latronico}, {Lemoine-Goumard}, {Longo}, {Loparco}, {Lott}, {Lovellette},
  {Lubrano}, {Mayer}, {Mazziotta}, {McEnery}, {Michelson}, {Mitthumsiri},
  {Mizuno}, {Monte}, {Monzani}, {Morselli}, {Moskalenko}, {Murgia}, {Nemmen},
  {Nuss}, {Ohsugi}, {Okumura}, {Omodei}, {Orienti}, {Orlando}, {Ormes},
  {Paneque}, {Panetta}, {Perkins}, {Pesce-Rollins}, {Piron}, {Pivato},
  {Porter}, {Rain{\`o}}, {Rando}, {Razzano}, {Reimer}, {Reimer}, {Romoli},
  {Roth}, {S{\'a}nchez-Conde}, {Scargle}, {Schulz}, {Sgr{\`o}}, {Siskind},
  {Spandre}, {Spinelli}, {Suson}, {Takahashi}, {Takeuchi}, {Thayer}, {Thayer},
  {Thompson}, {Tibaldo}, {Tinivella}, {Torres}, {Tosti}, {Troja}, {Tronconi},
  {Usher}, {Vandenbroucke}, {Vasileiou}, {Vianello}, {Vitale}, {Winer}, {Wood},
  {Wood}, \& {Yang}}]{FAVA2013}
{Ackermann}, M., {Ajello}, M., {Albert}, A., {et~al.} 2013{\natexlab{b}}, \apj,
  771, 57

\bibitem[{{Ackermann} {et~al.}(2013{\natexlab{c}}){Ackermann}, {Ajello},
  {Asano}, {Axelsson}, {Baldini}, {Ballet}, {Barbiellini}, {Bastieri},
  {Bechtol}, {Bellazzini}, {Bhat}, {Bissaldi}, {Bloom}, {Bonamente}, {Bonnell},
  {Bouvier}, {Brandt}, {Bregeon}, {Brigida}, {Bruel}, {Buehler}, {Burgess},
  {Buson}, {Byrne}, {Caliandro}, {Cameron}, {Caraveo}, {Cecchi}, {Charles},
  {Chaves}, {Chekhtman}, {Chiang}, {Chiaro}, {Ciprini}, {Claus},
  {Cohen-Tanugi}, {Connaughton}, {Conrad}, {Cutini}, {D'Ammando}, {de Angelis},
  {de Palma}, {Dermer}, {Desiante}, {Digel}, {Dingus}, {Di Venere}, {Drell},
  {Drlica-Wagner}, {Dubois}, {Favuzzi}, {Ferrara}, {Fitzpatrick}, {Foley},
  {Franckowiak}, {Fukazawa}, {Fusco}, {Gargano}, {Gasparrini}, {Gehrels},
  {Germani}, {Giglietto}, {Giommi}, {Giordano}, {Giroletti}, {Glanzman},
  {Godfrey}, {Goldstein}, {Granot}, {Grenier}, {Grove}, {Gruber}, {Guiriec},
  {Hadasch}, {Hanabata}, {Hayashida}, {Horan}, {Hou}, {Hughes}, {Inoue},
  {Jackson}, {Jogler}, {J{\'o}hannesson}, {Johnson}, {Johnson}, {Kamae},
  {Kataoka}, {Kawano}, {Kippen}, {Kn{\"o}dlseder}, {Kocevski}, {Kouveliotou},
  {Kuss}, {Lande}, {Larsson}, {Latronico}, {Lee}, {Longo}, {Loparco},
  {Lovellette}, {Lubrano}, {Massaro}, {Mayer}, {Mazziotta}, {McBreen},
  {McEnery}, {McGlynn}, {Michelson}, {Mizuno}, {Moiseev}, {Monte}, {Monzani},
  {Moretti}, {Morselli}, {Murgia}, {Nemmen}, {Nuss}, {Nymark}, {Ohno},
  {Ohsugi}, {Omodei}, {Orienti}, {Orlando}, {Paciesas}, {Paneque}, {Panetta},
  {Pelassa}, {Perkins}, {Pesce-Rollins}, {Piron}, {Pivato}, {Porter}, {Preece},
  {Racusin}, {Rain{\`o}}, {Rando}, {Rau}, {Razzano}, {Razzaque}, {Reimer},
  {Reimer}, {Reposeur}, {Ritz}, {Romoli}, {Roth}, {Ryde}, {Saz Parkinson},
  {Schalk}, {Sgr{\`o}}, {Siskind}, {Sonbas}, {Spandre}, {Spinelli}, {Suson},
  {Tajima}, {Takahashi}, {Takeuchi}, {Tanaka}, {Thayer}, {Thayer}, {Thompson},
  {Tibaldo}, {Tierney}, {Tinivella}, {Torres}, {Tosti}, {Troja}, {Tronconi},
  {Usher}, {Vandenbroucke}, {van der Horst}, {Vasileiou}, {Vianello}, {Vitale},
  {von Kienlin}, {Winer}, {Wood}, {Wood}, {Xiong}, \&
  {Yang}}]{2013ApJS..209...11A}
{Ackermann}, M., {Ajello}, M., {Asano}, K., {et~al.} 2013{\natexlab{c}}, \apjs,
  209, 11

\bibitem[{{Ackermann} {et~al.}(2014){Ackermann}, {Ajello}, {Asano}, {Atwood},
  {Axelsson}, {Baldini}, {Ballet}, {Barbiellini}, {Baring}, {Bastieri},
  {Bechtol}, {Bellazzini}, {Bissaldi}, {Bonamente}, {Bregeon}, {Brigida},
  {Bruel}, {Buehler}, {Burgess}, {Buson}, {Caliandro}, {Cameron}, {Caraveo},
  {Cecchi}, {Chaplin}, {Charles}, {Chekhtman}, {Cheung}, {Chiang}, {Chiaro},
  {Ciprini}, {Claus}, {Cleveland}, {Cohen-Tanugi}, {Collazzi}, {Cominsky},
  {Connaughton}, {Conrad}, {Cutini}, {D'Ammando}, {de Angelis}, {DeKlotz}, {de
  Palma}, {Dermer}, {Desiante}, {Diekmann}, {Di Venere}, {Drell},
  {Drlica-Wagner}, {Favuzzi}, {Fegan}, {Ferrara}, {Finke}, {Fitzpatrick},
  {Focke}, {Franckowiak}, {Fukazawa}, {Funk}, {Fusco}, {Gargano}, {Gehrels},
  {Germani}, {Gibby}, {Giglietto}, {Giles}, {Giordano}, {Giroletti}, {Godfrey},
  {Granot}, {Grenier}, {Grove}, {Gruber}, {Guiriec}, {Hadasch}, {Hanabata},
  {Harding}, {Hayashida}, {Hays}, {Horan}, {Hughes}, {Inoue}, {Jogler},
  {J{\'o}hannesson}, {Johnson}, {Kawano}, {Kn{\"o}dlseder}, {Kocevski}, {Kuss},
  {Lande}, {Larsson}, {Latronico}, {Longo}, {Loparco}, {Lovellette}, {Lubrano},
  {Mayer}, {Mazziotta}, {McEnery}, {Michelson}, {Mizuno}, {Moiseev}, {Monzani},
  {Moretti}, {Morselli}, {Moskalenko}, {Murgia}, {Nemmen}, {Nuss}, {Ohno},
  {Ohsugi}, {Okumura}, {Omodei}, {Orienti}, {Paneque}, {Pelassa}, {Perkins},
  {Pesce-Rollins}, {Petrosian}, {Piron}, {Pivato}, {Porter}, {Racusin},
  {Rain{\`o}}, {Rando}, {Razzano}, {Razzaque}, {Reimer}, {Reimer}, {Ritz},
  {Roth}, {Ryde}, {Sartori}, {Parkinson}, {Scargle}, {Schulz}, {Sgr{\`o}},
  {Siskind}, {Sonbas}, {Spandre}, {Spinelli}, {Tajima}, {Takahashi}, {Thayer},
  {Thayer}, {Thompson}, {Tibaldo}, {Tinivella}, {Torres}, {Tosti}, {Troja},
  {Usher}, {Vandenbroucke}, {Vasileiou}, {Vianello}, {Vitale}, {Winer}, {Wood},
  {Yamazaki}, {Younes}, {Yu}, {Zhu}, {Bhat}, {Briggs}, {Byrne}, {Foley},
  {Goldstein}, {Jenke}, {Kippen}, {Kouveliotou}, {McBreen}, {Meegan},
  {Paciesas}, {Preece}, {Rau}, {Tierney}, {van der Horst}, {von Kienlin},
  {Wilson-Hodge}, {Xiong}, {Cusumano}, {La Parola}, \&
  {Cummings}}]{Ackermann2014}
---. 2014, Science, 343, 42

\bibitem[{{Ackermann} {et~al.}(2016){Ackermann}, {Ajello}, {Albert},
  {Anderson}, {Arimoto}, {Atwood}, {Axelsson}, {Baldini}, {Ballet},
  {Barbiellini}, {Baring}, {Bastieri}, {Becerra Gonzalez}, {Bellazzini},
  {Bissaldi}, {Blandford}, {Bloom}, {Bonino}, {Bottacini}, {Brandt}, {Bregeon},
  {Britto}, {Bruel}, {Buehler}, {Burnett}, {Buson}, {Caliandro}, {Cameron},
  {Caputo}, {Caragiulo}, {Caraveo}, {Casandjian}, {Cavazzuti}, {Charles},
  {Chekhtman}, {Chiang}, {Chiaro}, {Ciprini}, {Cohen-Tanugi}, {Cominsky},
  {Condon}, {Costanza}, {Cuoco}, {Cutini}, {D'Ammando}, {de Palma}, {Desiante},
  {Digel}, {Di Lalla}, {Di Mauro}, {Di Venere}, {Dom{\'{\i}}nguez}, {Drell},
  {Dubois}, {Dumora}, {Favuzzi}, {Fegan}, {Ferrara}, {Franckowiak}, {Fukazawa},
  {Funk}, {Fusco}, {Gargano}, {Gasparrini}, {Gehrels}, {Giglietto}, {Giomi},
  {Giommi}, {Giordano}, {Giroletti}, {Glanzman}, {Godfrey}, {Gomez-Vargas},
  {Granot}, {Green}, {Grenier}, {Grondin}, {Grove}, {Guillemot}, {Guiriec},
  {Hadasch}, {Harding}, {Hays}, {Hewitt}, {Hill}, {Horan}, {Jogler},
  {J{\'o}hannesson}, {Kamae}, {Kensei}, {Kocevski}, {Kuss}, {La Mura},
  {Larsson}, {Latronico}, {Lemoine-Goumard}, {Li}, {Li}, {Longo}, {Loparco},
  {Lovellette}, {Lubrano}, {Madejski}, {Magill}, {Maldera}, {Manfreda},
  {Marelli}, {Mayer}, {Mazziotta}, {McEnery}, {Meyer}, {Michelson}, {Mirabal},
  {Mizuno}, {Moiseev}, {Monzani}, {Moretti}, {Morselli}, {Moskalenko},
  {Murgia}, {Negro}, {Nuss}, {Ohsugi}, {Omodei}, {Orienti}, {Orlando}, {Ormes},
  {Paneque}, {Perkins}, {Pesce-Rollins}, {Piron}, {Pivato}, {Porter},
  {Racusin}, {Rain{\`o}}, {Rando}, {Razzaque}, {Reimer}, {Reimer}, {Reposeur},
  {Ritz}, {Rochester}, {Romani}, {Saz Parkinson}, {Sgr{\`o}}, {Simone},
  {Siskind}, {Smith}, {Spada}, {Spandre}, {Spinelli}, {Suson}, {Tajima},
  {Thayer}, {Thayer}, {Thompson}, {Tibaldo}, {Torres}, {Troja}, {Uchiyama},
  {Venters}, {Vianello}, {Wood}, {Wood}, {Zaharijas}, {Zhu}, \&
  {Zimmer}}]{Ackermann2016}
{Ackermann}, M., {Ajello}, M., {Albert}, A., {et~al.} 2016, \apjl, 823, L2

\bibitem[{{Atwood} {et~al.}(2009){Atwood}, {Abdo}, {Ackermann}, {Althouse},
  {Anderson}, {Axelsson}, {Baldini}, {Ballet}, {Band}, {Barbiellini}, \&
  et~al.}]{Atwood09}
{Atwood}, W.~B., {Abdo}, A.~A., {Ackermann}, M., {et~al.} 2009, \apj, 697, 1071

\bibitem[{{Berger}(2014)}]{Berger2014}
{Berger}, E. 2014, \araa, 52, 43

\bibitem[{{Berger} {et~al.}(2013){Berger}, {Fong}, \& {Chornock}}]{Berger2013}
{Berger}, E., {Fong}, W., \& {Chornock}, R. 2013, \apjl, 774, L23

\bibitem[{Bhat {et~al.}(2016)Bhat, Meegan, von Kienlin, Paciesas, Briggs,
  Burgess, Burns, Chaplin, Cleveland, Collazzi, Connaughton, Diekmann,
  Fitzpatrick, Gibby, Giles, Goldstein, Greiner, Jenke, Kippen, Kouveliotou,
  Mailyan, McBreen, Pelassa, Preece, Roberts, Sparke, Stanbro, Veres,
  Wilson-Hodge, Xiong, Younes, Yu, \& Zhang}]{GBMBurstCatalog_6Years}
Bhat, P.~N., Meegan, C.~A., von Kienlin, A., {et~al.} 2016, ApJS, 223, 28

\bibitem[{{Bilicki} {et~al.}(2014){Bilicki}, {Jarrett}, {Peacock}, {Cluver}, \&
  {Steward}}]{2MPZ}
{Bilicki}, M., {Jarrett}, T.~H., {Peacock}, J.~A., {Cluver}, M.~E., \&
  {Steward}, L. 2014, \apjs, 210, 9

\bibitem[{{Chiang} {et~al.}(2007){Chiang}, {Carson}, \&
  {Focke}}]{2007AIPC..921..544C}
{Chiang}, J., {Carson}, J., \& {Focke}, W. 2007, in American Institute of
  Physics Conference Series, Vol. 921, The First GLAST Symposium, ed.
  S.~{Ritz}, P.~{Michelson}, \& C.~A. {Meegan}, 544--545

\bibitem[{{Connaughton} {et~al.}(2017)}]{GRB170817A_Discovery}
{Connaughton}, V., {et~al.} 2017, Gamma Ray Coordinates Network Circular, 21506

\bibitem[{{Coulter} {et~al.}(2017{\natexlab{a}}){Coulter}, {Kilpatrick},
  {Siebert}, {Foley}, {Shappee}, {Drout}, \& {Piro}}]{GCN21529}
{Coulter}, D.~A., {Kilpatrick}, C.~D., {Siebert}, M.~R., {et~al.}
  2017{\natexlab{a}}, GRB Coordinates Network, 21529

\bibitem[{{Coulter} {et~al.}(2017{\natexlab{b}})}]{Coulter2017}
{Coulter}, D.~A., {et~al.} 2017{\natexlab{b}}, Science,
  doi:10.1126/science.aap9811

\bibitem[{{D'Avanzo}(2015)}]{Avanzo:2015}
{D'Avanzo}, P. 2015, Journal of High Energy Astrophysics, 7, 73

\bibitem[{{D'Avanzo} {et~al.}(2009){D'Avanzo}, {Malesani}, {Covino},
  {Piranomonte}, {Grazian}, {Fugazza}, {Margutti}, {D'Elia}, {Antonelli},
  {Campana}, {Chincarini}, {Della Valle}, {Fiore}, {Goldoni}, {Mao}, {Perna},
  {Salvaterra}, {Stella}, {Stratta}, \& {Tagliaferri}}]{2009A&A...498..711D}
{D'Avanzo}, P., {Malesani}, D., {Covino}, S., {et~al.} 2009, \aap, 498, 711

\bibitem[{Dezalay {et~al.}(1991)Dezalay, Barat, Talon, Sunyaev, Terekhov, \&
  Kuznetsov}]{Dezalay1991}
Dezalay, J., Barat, C., Talon, R., {et~al.} 1991, AIP Conference Proceedings,
  265, 304

\bibitem[{{Eichler} {et~al.}(1989){Eichler}, {Livio}, {Piran}, \&
  {Schramm}}]{Eichler1989}
{Eichler}, D., {Livio}, M., {Piran}, T., \& {Schramm}, D.~N. 1989, \nat, 340,
  126

\bibitem[{{Fong} \& {Berger}(2013)}]{Fong2013}
{Fong}, W., \& {Berger}, E. 2013, \apj, 776, 18

\bibitem[{{Fox} {et~al.}(2005){Fox}, {Frail}, {Price}, {Kulkarni}, {Berger},
  {Piran}, {Soderberg}, {Cenko}, {Cameron}, {Gal-Yam}, {Kasliwal}, {Moon},
  {Harrison}, {Nakar}, {Schmidt}, {Penprase}, {Chevalier}, {Kumar}, {Roth},
  {Watson}, {Lee}, {Shectman}, {Phillips}, {Roth}, {McCarthy}, {Rauch},
  {Cowie}, {Peterson}, {Rich}, {Kawai}, {Aoki}, {Kosugi}, {Totani}, {Park},
  {MacFadyen}, \& {Hurley}}]{Fox2005}
{Fox}, D.~B., {Frail}, D.~A., {Price}, P.~A., {et~al.} 2005, \nat, 437, 845

\bibitem[{{Goldstein} {et~al.}(2017)}]{Goldstein2017}
{Goldstein}, A., {et~al.} 2017, \apjl, 848

\bibitem[{{Granot} {et~al.}(2002){Granot}, {Panaitescu}, {Kumar}, \&
  {Woosley}}]{Granot:2002}
{Granot}, J., {Panaitescu}, A., {Kumar}, P., \& {Woosley}, S.~E. 2002, \apjl,
  570, L61

\bibitem[{{Hjorth} {et~al.}(2005{\natexlab{a}}){Hjorth}, {Watson}, {Fynbo},
  {Price}, {Jensen}, {J{\o}rgensen}, {Kubas}, {Gorosabel}, {Jakobsson},
  {Sollerman}, {Pedersen}, \& {Kouveliotou}}]{Hjorth2005a}
{Hjorth}, J., {Watson}, D., {Fynbo}, J.~P.~U., {et~al.} 2005{\natexlab{a}},
  \nat, 437, 859

\bibitem[{{Hjorth} {et~al.}(2005{\natexlab{b}}){Hjorth}, {Sollerman},
  {Gorosabel}, {Granot}, {Klose}, {Kouveliotou}, {Melinder}, {Ramirez-Ruiz},
  {Starling}, {Thomsen}, {Andersen}, {Fynbo}, {Jensen}, {Vreeswijk}, {Castro
  Cer{\'o}n}, {Jakobsson}, {Levan}, {Pedersen}, {Rhoads}, {Tanvir}, {Watson},
  \& {Wijers}}]{Hjorth2005b}
{Hjorth}, J., {Sollerman}, J., {Gorosabel}, J., {et~al.} 2005{\natexlab{b}},
  \apjl, 630, L117

\bibitem[{{Jin} {et~al.}(2016){Jin}, {Hotokezaka}, {Li}, {Tanaka}, {D'Avanzo},
  {Fan}, {Covino}, {Wei}, \& {Piran}}]{Jin2016NatCo...712898J}
{Jin}, Z.-P., {Hotokezaka}, K., {Li}, X., {et~al.} 2016, Nature Communications,
  7, 12898

\bibitem[{{Kashyap} {et~al.}(2010){Kashyap}, {van Dyk}, {Connors}, {Freeman},
  {Siemiginowska}, {Xu}, \& {Zezas}}]{2010ApJ...719..900K}
{Kashyap}, V.~L., {van Dyk}, D.~A., {Connors}, A., {et~al.} 2010, \apj, 719,
  900

\bibitem[{{Kobayashi} \& {M{\'e}sz{\'a}ros}(2003)}]{Kobayashi2003}
{Kobayashi}, S., \& {M{\'e}sz{\'a}ros}, P. 2003, \apj, 589, 861

\bibitem[{{Kocevski} {et~al.}(2010){Kocevski}, {Th{\"o}ne}, {Ramirez-Ruiz},
  {Bloom}, {Granot}, {Butler}, {Perley}, {Modjaz}, {Lee}, {Cobb}, {Levan},
  {Tanvir}, \& {Covino}}]{kocevski2010}
{Kocevski}, D., {Th{\"o}ne}, C.~C., {Ramirez-Ruiz}, E., {et~al.} 2010, \mnras,
  404, 963

\bibitem[{{Kouveliotou} {et~al.}(1993){Kouveliotou}, {Meegan}, {Fishman},
  {Bhat}, {Briggs}, {Koshut}, {Paciesas}, \& {Pendleton}}]{Kouveliotou1993}
{Kouveliotou}, C., {Meegan}, C.~A., {Fishman}, G.~J., {et~al.} 1993, \apjl,
  413, L101

\bibitem[{{Kouveliotou} {et~al.}(2012){Kouveliotou}, {Wijers}, \&
  {Woosley}}]{2012grb..book.....K}
{Kouveliotou}, C., {Wijers}, R.~A.~M.~J., \& {Woosley}, S. 2012, {Gamma-ray
  Bursts}

\bibitem[{{Kouveliotou} {et~al.}(2013){Kouveliotou}, {Granot}, {Racusin},
  {Bellm}, {Vianello}, {Oates}, {Fryer}, {Boggs}, {Christensen}, {Craig},
  {Dermer}, {Gehrels}, {Hailey}, {Harrison}, {Melandri}, {McEnery}, {Mundell},
  {Stern}, {Tagliaferri}, \& {Zhang}}]{Kouveliotou2013}
{Kouveliotou}, C., {Granot}, J., {Racusin}, J.~L., {et~al.} 2013, \apjl, 779,
  L1

\bibitem[{{Kumar} \& {Barniol Duran}(2009)}]{Kumar2009}
{Kumar}, P., \& {Barniol Duran}, R. 2009, \mnras, 400, L75

\bibitem[{{Meegan} {et~al.}(2009){Meegan}, {Lichti}, {Bhat}, {Bissaldi},
  {Briggs}, {Connaughton}, {Diehl}, {Fishman}, {Greiner}, {Hoover}, {van der
  Horst}, {von Kienlin}, {Kippen}, {Kouveliotou}, {McBreen}, {Paciesas},
  {Preece}, {Steinle}, {Wallace}, {Wilson}, \& {Wilson-Hodge}}]{Meegan2009}
{Meegan}, C., {Lichti}, G., {Bhat}, P.~N., {et~al.} 2009, \apj, 702, 791

\bibitem[{Metzger {et~al.}(2010)Metzger, Martinez-Pinedo, Darbha, Quataert,
  Arcones, Kasen, Thomas, Nugent, Panov, \& Zinner}]{Metzger2010}
Metzger, B.~D., Martinez-Pinedo, G., Darbha, S., {et~al.} 2010, Mon. Not. Roy.
  Astron. Soc., 406, 2650

\bibitem[{Mooley {et~al.}(2017)}]{GCN21814}
Mooley, K.~P., {et~al.} 2017, GCN, 21814, 1

\bibitem[{{Narayan} {et~al.}(1992){Narayan}, {Paczynski}, \&
  {Piran}}]{Narayan1992}
{Narayan}, R., {Paczynski}, B., \& {Piran}, T. 1992, \apjl, 395, L83

\bibitem[{{Norris} {et~al.}(1984){Norris}, {Cline}, {Desai}, \&
  {Teegarden}}]{Norris1984}
{Norris}, J.~P., {Cline}, T.~L., {Desai}, U.~D., \& {Teegarden}, B.~J. 1984,
  \nat, 308, 434

\bibitem[{{Paczynski}(1986)}]{Paczynski1986}
{Paczynski}, B. 1986, \apjl, 308, L43

\bibitem[{{Paczynski}(1991)}]{Paczynski1991}
---. 1991, \actaa, 41, 257

\bibitem[{{Patricelli} {et~al.}(2016){Patricelli}, {Razzano}, {Cella},
  {Fidecaro}, {Pian}, {Branchesi}, \& {Stamerra}}]{2016JCAP...11..056P}
{Patricelli}, B., {Razzano}, M., {Cella}, G., {et~al.} 2016, \jcap, 11, 056

\bibitem[{{Planck Collaboration} {et~al.}(2016){Planck Collaboration}, {Ade},
  {Aghanim}, {Arnaud}, {Ashdown}, {Aumont}, {Baccigalupi}, {Banday},
  {Barreiro}, \& {Bartlett}}]{2016A&A...594A..13P}
{Planck Collaboration}, {Ade}, P.~A.~R., {Aghanim}, N., {et~al.} 2016, \aap,
  594, A13

\bibitem[{{Racusin} {et~al.}(2017){Racusin}, {Burns}, {Goldstein},
  {Connaughton}, {Wilson-Hodge}, {Jenke}, {Blackburn}, {Briggs}, {Broida},
  {Camp}, {Christensen}, {Hui}, {Littenberg}, {Shawhan}, {Singer}, {Veitch},
  {Bhat}, {Cleveland}, {Fitzpatrick}, {Gibby}, {von Kienlin}, {McBreen},
  {Mailyan}, {Meegan}, {Paciesas}, {Preece}, {Roberts}, {Stanbro}, {Veres},
  {Zhang}, {Fermi LAT Collaboration}, {Ackermann}, {Albert}, {Atwood},
  {Axelsson}, {Baldini}, {Ballet}, {Barbiellini}, {Baring}, {Bastieri},
  {Bellazzini}, {Bissaldi}, {Blandford}, {Bloom}, {Bonino}, {Bregeon}, {Bruel},
  {Buson}, {Caliandro}, {Cameron}, {Caputo}, {Caragiulo}, {Caraveo},
  {Cavazzuti}, {Charles}, {Chiang}, {Ciprini}, {Costanza}, {Cuoco}, {Cutini},
  {D'Ammando}, {de Palma}, {Desiante}, {Digel}, {Di Lalla}, {Di Mauro}, {Di
  Venere}, {Drell}, {Favuzzi}, {Ferrara}, {Focke}, {Fukazawa}, {Funk}, {Fusco},
  {Gargano}, {Gasparrini}, {Giglietto}, {Gill}, {Giroletti}, {Glanzman},
  {Granot}, {Green}, {Grove}, {Guillemot}, {Guiriec}, {Harding}, {Jogler},
  {J{\'o}hannesson}, {Kamae}, {Kensei}, {Kocevski}, {Kuss}, {Larsson},
  {Latronico}, {Li}, {Longo}, {Loparco}, {Lubrano}, {Magill}, {Maldera},
  {Malyshev}, {Mazziotta}, {McEnery}, {Michelson}, {Mizuno}, {Monzani},
  {Morselli}, {Moskalenko}, {Negro}, {Nuss}, {Omodei}, {Orienti}, {Orlando},
  {Ormes}, {Paneque}, {Perkins}, {Pesce-Rollins}, {Piron}, {Pivato}, {Porter},
  {Principe}, {Rain{\`o}}, {Rando}, {Razzano}, {Razzaque}, {Reimer}, {Reimer},
  {Saz Parkinson}, {Scargle}, {Sgr{\`o}}, {Simone}, {Siskind}, {Smith},
  {Spada}, {Spinelli}, {Suson}, {Tajima}, {Thayer}, {Torres}, {Troja},
  {Uchiyama}, {Vianello}, {Wood}, \& {Wood}}]{Racusin2017}
{Racusin}, J.~L., {Burns}, E., {Goldstein}, A., {et~al.} 2017, \apj, 835, 82

\bibitem[{{Razzaque}(2010)}]{Razzaque2010}
{Razzaque}, S. 2010, \apjl, 724, L109

\bibitem[{{Savchenko} {et~al.}(2017{\natexlab{a}})}]{GCN21507}
{Savchenko}, V., {et~al.} 2017{\natexlab{a}}, GCN, 21507, 1

\bibitem[{{Savchenko} {et~al.}(2017{\natexlab{b}}){Savchenko}, {Ferrigno},
  {Kuulkers}, {Bazzano}, {Bozzo}, {Brandt}, {Courvoisier}, {Diehl}, {Hanlon},
  {Laurent}, {Mereghetti}, {Roques}, \& {Ubertini}}]{Savcheno2017}
{Savchenko}, V., {Ferrigno}, C., {Kuulkers}, E., {et~al.} 2017{\natexlab{b}},
  ApJL, in press

\bibitem[{{Soderberg} {et~al.}(2006){Soderberg}, {Berger}, {Kasliwal}, {Frail},
  {Price}, {Schmidt}, {Kulkarni}, {Fox}, {Cenko}, {Gal-Yam}, {Nakar}, \&
  {Roth}}]{Soderberg2006}
{Soderberg}, A.~M., {Berger}, E., {Kasliwal}, M., {et~al.} 2006, \apj, 650, 261

\bibitem[{{Tanvir} {et~al.}(2013){Tanvir}, {Levan}, {Fruchter}, {Hjorth},
  {Hounsell}, {Wiersema}, \& {Tunnicliffe}}]{tlf+2013}
{Tanvir}, N.~R., {Levan}, A.~J., {Fruchter}, A.~S., {et~al.} 2013, \nat, 500,
  547

\bibitem[{{Taylor} {et~al.}(1979){Taylor}, {Fowler}, \&
  {McCulloch}}]{1979Natur.277..437T}
{Taylor}, J.~H., {Fowler}, L.~A., \& {McCulloch}, P.~M. 1979, \nat, 277, 437

\bibitem[{{The {LIGO} Scientific Collaboration} {et~al.}(2017){The {LIGO}
  Scientific Collaboration}, {The Virgo Collaboration},
  {et~al.}}]{gcnLVCGW170817_LAL}
{The {LIGO} Scientific Collaboration}, {The Virgo Collaboration}, {et~al.}
  2017, GCN, 21527, 1

\bibitem[{{Troja} {et~al.}(2008){Troja}, {King}, {O'Brien}, {Lyons}, \&
  {Cusumano}}]{2008MNRAS.385L..10T}
{Troja}, E., {King}, A.~R., {O'Brien}, P.~T., {Lyons}, N., \& {Cusumano}, G.
  2008, \mnras, 385, L10

\bibitem[{{Troja} {et~al.}(2016){Troja}, {Sakamoto}, {Cenko}, {Lien},
  {Gehrels}, {Castro-Tirado}, {Ricci}, {Capone}, {Toy}, {Kutyrev}, {Kawai},
  {Cucchiara}, {Fruchter}, {Gorosabel}, {Jeong}, {Levan}, {Perley},
  {Sanchez-Ramirez}, {Tanvir}, \& {Veilleux}}]{2016ApJ...827..102T}
{Troja}, E., {Sakamoto}, T., {Cenko}, S.~B., {et~al.} 2016, \apj, 827, 102

\bibitem[{Troja {et~al.}(2017)}]{GCN21787}
Troja, E., {et~al.} 2017, GCN, 21787, 1

\bibitem[{{Troja} {et~al.}(2017){Troja}, {Piro}, {van Eerten}, {Wollaeger},
  {Im}, {Fox}, {Butler}, {Cenko}, {Sakamoto}, {Fyer}, {Ricci}, {Lien}, \&
  {others}}]{Troja:2017}
{Troja}, E., {Piro}, L., {van Eerten}, H., {et~al.} 2017, \nat,
  doi:10.1038/nature24290

\bibitem[{{van Eerten} \& {MacFadyen}(2012)}]{vanEerten2012}
{van Eerten}, H.~J., \& {MacFadyen}, A.~I. 2012, \apj, 751, 155

\bibitem[{Veitch {et~al.}(2015)Veitch, Raymond, Farr, Farr, Graff, Vitale,
  Aylott, Blackburn, Christensen, Coughlin, \& et~al.}]{veitch2015parameter}
Veitch, J., Raymond, V., Farr, B., {et~al.} 2015, Physical Review D, 91, 042003

\bibitem[{{Vianello} {et~al.}(2017){Vianello}, {Omodei}, {Chiang}, \&
  {Digel}}]{Vianello2017}
{Vianello}, G., {Omodei}, N., {Chiang}, J., \& {Digel}, S. 2017, \apjl, 841,
  L16

\bibitem[{{Vianello} {et~al.}(2015){Vianello}, {Omodei}, \& {Fermi/LAT
  collaboration}}]{Vianello15}
{Vianello}, G., {Omodei}, N., \& {Fermi/LAT collaboration}. 2015, ArXiv
  e-prints, arXiv:1502.03122

\bibitem[{{Virgo Collaboration}(2009)}]{AdvVirgo}
{Virgo Collaboration}. 2009, Advanced Virgo Baseline Design, Virgo Technical
  Report VIR-0027A-09,
  \url{https://tds.ego-gw.it/itf/tds/file.php?callFile=VIR-0027A-09.pdf}

\bibitem[{{Yang} {et~al.}(2015){Yang}, {Jin}, {Li}, {Covino}, {Zheng},
  {Hotokezaka}, {Fan}, {Piran}, \& {Wei}}]{Yang2015NatCo...6E7323Y}
{Yang}, B., {Jin}, Z.-P., {Li}, X., {et~al.} 2015, Nature Communications, 6,
  7323

\end{thebibliography}

\end{document}